\documentclass[usenatbib]{mn2e}
\usepackage{bm}
\usepackage{morefloats}
\usepackage{hyperref}
\usepackage{tabularx}
\usepackage{graphicx}
\usepackage{multirow}
\usepackage{pdflscape}
\usepackage{amsmath,amssymb}

\newcommand {\gaia}{\textit{Gaia }}
\newcommand {\hst}{\textit{HST }}

\title[Stellar mass segregation in GCs and UFDs]
{Stellar mass segregation as separating classifier between globular clusters and ultra-faint dwarf galaxies}
\author[Baumgardt, Faller, Meinhold, McGovern-Greco, Hilker]{H. Baumgardt$^{1}$\thanks{E-mail:
h.baumgardt@uq.edu.au}, J. Faller$^{1}$, N. Meinhold$^{1}$, C. McGovern-Greco$^{1}$, M. Hilker$^{2}$ \\
$^{1}$ School of Mathematics and Physics, The University of Queensland, St. Lucia, QLD 4072, Australia \\
$^{2}$ European Southern Observatory, Karl-Schwarzschild-Str. 2, 85748 Garching, Germany\\
}

\begin{document}

\date{Accepted 2021 xx xx. Received 2021 xx xx; in original form 2021 xx xx}

\pagerange{\pageref{firstpage}--\pageref{lastpage}} \pubyear{201x}

\maketitle

\label{firstpage}

\begin{abstract}
We have determined the amount of stellar mass segregation in over 50 globular clusters and ultra-faint dwarf galaxy candidates based on deep \hst and ground-based photometry. We find
that the amount of mass segregation in globular clusters is strongly correlated with their relaxation time and that all clusters with relaxation times of the order 
of their ages or longer have little to no mass segregation. For each cluster, the amount of mass segregation seen is fully compatible with the amount expected by dynamical evolution from initially
unsegregated clusters, showing that globular clusters formed without primordial mass segregation among their low-mass stars.
Ultra-faint dwarf galaxy candidates 
split into two groups, star clusters which follow the same trend between relaxation time and amount of mass segregation as globular clusters and dark-matter dominated dwarf
galaxies that are unsegregated despite having relaxation times smaller than a Hubble time. Stellar abundance and velocity dispersion data, where available, confirm our classification. After 
classification of the ultra-faint dwarf galaxy candidates, we find that outer halo star clusters have average densities inside their half-light radii of 
0.03~M$_\odot$/pc$^3 \lesssim \rho_h \lesssim$ 1~M$_\odot$/pc$^3$, while  dwarf galaxies have stellar densities of 0.001 M$_\odot$/pc$^3 \lesssim \rho_h \lesssim$ 0.03~M$_\odot$/pc$^3$.  
The reason for this separation in density is most likely a combination of the initial conditions by which the systems formed 
and the requirement to withstand external tidal forces.
\end{abstract}


\section{Introduction} \label{sec:intro}

Dwarf galaxies have in recent years become useful probes to test the properties of dark matter. Determining their absolute numbers and mass function is an important test for our
understanding of galaxy formation on small scales and allows to
put constraints on the mass of the dark matter particle \citep{bodeetal2001,jethwaetal2018}. Furthermore their central densities can provide clues to the behavior of the dark matter
particle on small scales \citep[e.g.][]{safarzadehspergel2020,hayashietal2021}. Dwarf galaxies are also useful tools to enhance our understanding of galaxy formation. Especially the faintest dwarf galaxies 
provide vital insights into the star formation process in the early universe, since their low
stellar masses and low metallicities mean that they formed at high redshift and have likely the simplest enrichment history of all galaxies. This makes them pristine relics of the early universe \citep{blandhawthornetal2015,romanoetal2019}.

The number of known dwarf galaxy candidates has increased dramatically in the last fifteen years thanks in large part to wide area digital sky surveys \citep[see e.g. Fig.~1 in ][]{simon2019}. Most of the newly 
discovered dwarf galaxies
are ultra-faint dwarf galaxies (UFDs) with luminosities $M_V>-7$ and sizes of a few tens of pc. The location of these UFDs in a size vs. mass plane overlaps with that of faint globular clusters.  It is therefore important
to obtain additional information to be able to separate UFDs from halo star clusters. Two common ways to identify UFDs are either through their kinematics or by measuring the stellar abundances of their member 
stars. While the kinematics of globular clusters can be well explained based on the visible stars alone \citep[e.g.][]{baumgardthilker2018}, UFDs are found to have velocity dispersions significantly larger than those 
expected based solely on their visible stars. The measured velocity dispersions of UFDs are so large that their total masses must be orders of magnitude higher than their stellar masses \citep[e.g.][]{simongeha2007}.
Furthermore the faintest dwarf galaxies have generally low iron abundances (down to about [Fe/H]$\sim -3.0$), while the more massive dwarf galaxies show large abundance spreads 
between their member stars, indicating prolonged star formation histories.  In addition, low-mass UFDs have been found to be enriched in $r$-process elements \citep{jietal2016}, indicating that their stars were enriched by
single core-collapse supernova events. Globular clusters on the other hand have a large range of metallicities, their internal iron abundance spreads are small and limited to the most massive clusters,
and abundance spreads in other elements occur mainly among the light elements \citep{gratton2004,carrettaetal2010b,roedigeretal2014}.

Both the abundance as well as the velocity dispersion test become increasingly harder to apply for fainter UFDs since  the faintest detected UFDs contain few if any giant stars for which high-resolution spectra can be obtained.
Measuring velocity dispersions from only a few stars is however subject to significant uncertainties: Apart from
the fact that the error bars reach a sizeable fraction of the measured dispersion itself, the measured dispersion also suffers from contamination by field stars and stellar binaries \citep[e.g.][]{kirbyetal2015b,kirbyetal2017}. 
Furthermore
a low metallicity alone is not a proof for a stellar system being a dwarf galaxy, since globular clusters with metallicities as low as [Fe/H]=-2.5 and [Fe/H]=-2.9 have been found in the Milky Way and
M31 respectively \citep{simpson2018,larsenetal2020}. Such metallicities are similar to the ones found for the most metal-poor dwarf galaxies like Bo{\"o}tes II \citep{kochrich2014}.

It is therefore important to obtain further information that can help to distinguish dwarf galaxies from star clusters among the known ultra-faint dwarf galaxy candidates in the halo of the Milky Way. One such possibility is to look for stellar mass segregation \citep[e.g.][]{longeardetal2018,connetal2018b}:
Due to energy equipartition, high mass stars are expected to sink into the centers
of star clusters while low-mass stars are pushed towards the outskirts of the clusters. Energy equipartition takes typically a few relaxation times to be established, old halo star clusters and globular clusters with relaxation
times less than a few Gyr should therefore 
be mass segregated. The presence of a significant amount of dark matter prolongs the timescale for mass segregation significantly, hence most low-mass dwarf galaxy candidates are predicted to be mass
segregated only if free of dark matter, i.e. if they are star clusters. If they contain significant amounts of dark matter, their relaxation times are larger than a Hubble time and they should
not be mass segregated.

Our paper is organized as follows: In sec. 2 we discuss the analysis of the \hst and ground based photometry and in sec.~3 we present our determination of the amount of mass segregation. In sec.~4 we discuss
the determination of mass segregation in a number of $N$-body simulations that we use to test the origin of the observed mass segregation. In sec.~5 we discuss our results of the amount of mass segregation in the
different systems and in sec.~6 we present our conclusions. 

\section{Photometry}
\label{sec:photometry}

We take the input list of globular clusters from \citet{baumgardtetal2019a}, concentrating on outer halo globular clusters that cover a similar range of Galactocentric distances
as the ultra-faint dwarf galaxy candidates. From the catalogue of \citet{baumgardtetal2019a} we exclude all clusters that lack deep enough photometry to measure the
distribution of low-mass stars. We also exclude clusters like Pal~2 that have strong and variable reddening or a high stellar background density. 
Finally, we remove clusters with high central surface densities in which the distribution of faint stars can't be measured in the centres. This criterion excludes most high-mass halo
clusters like M~54. 
We finally require that the existing photometry must cover the cluster from the centre out to at least 2.5 half-light radii. This criterion excludes most nearby clusters 
for which a single HST field of view is too small to cover a large enough part of the cluster. Applying all criteria, we end up with 28 globular clusters, having relaxation times
from 0.1 to 50 Gyrs and cluster masses from $7.5 \cdot 10^2$ M$_\odot$ to $3.9 \cdot 10^6$ M$_\odot$. Except for M~4 and $\omega$~Cen, all clusters are at Galactocentric distances larger than 10~kpc.
17 of these clusters have published photometry and we list the clusters and the sources of their photometry in Table~1.
\begin{table}
\caption{Published literature photometry used in the present work}
\begin{tabular}{l@{$\;$}l@{$\;$}l}
\hline
Star Cluster/ & \multirow{2}{*}{Telescope/Camera/Filter} & \multirow{2}{*}{Literature source} \\
UFD &  & \\
\hline
& & \\[-0.20cm]
Balbinot 1 & CFHT/MegaCam g/r &\citet{balbinotetal2013} \\[+0.10cm]
Kim 1      & CFHT/MegaCam g/r & \citet{munozetal2018c}  \\[+0.10cm]
Kim 2      & CFHT/MegaCam g/r & \citet{munozetal2018c}  \\[+0.10cm]
Koposov 1  & CFHT/MegaCam g/r & \citet{munozetal2018c} \\[+0.10cm]
Koposov 2  & CFHT/MegaCam g/r & \citet{munozetal2018c} \\[+0.10cm]
NGC 1261   & HST/ACS F606W/F814W & \citet{sarajedinietal2007} \\[+0.10cm]
NGC 5053   & HST/ACS F606W/F814W & \citet{sarajedinietal2007} \\
           & HST/ACS F475W/F814W & \citet{simionietal2018} \\[+0.10cm]
NGC 5466   & HST/ACS F606W/F814W & \citet{sarajedinietal2007} \\
           & HST/ACS F475W/F814W & \citet{simionietal2018} \\[+0.10cm]
NGC 5897   & HST/ACS F606W/F814W & \citet{sarajedinietal2007} \\
           & HST/ACS F475W/F814W & \citet{simionietal2018} \\[+0.10cm]
NGC 6121   & HST/ACS F606W/F814W & \citet{sarajedinietal2007} \\
           & Ground-based V/I    & \citet{stetsonetal2019} \\[+0.10cm]
NGC 6426   & HST/ACS F606W/F814W & Dotter et al. (2011)    \\[+0.10cm]
NGC 7006   & HST/ACS F606W/F814W & Dotter et al. (2011)    \\[+0.10cm]
NGC 7492   & Ground-based V/I    & \citet{stetsonetal2019} \\[+0.10cm]
Pal 1      & HST/ACS F606W/F814W & \citet{sarajedinietal2007} \\[+0.10cm]
Pal 5      & Ground-based V/I    & \citet{stetsonetal2019}    \\[+0.10cm]
Pal 12     & HST/ACS F606W/F814W & \citet{sarajedinietal2007} \\[+0.10cm]
Pal 15     & HST/ACS F606W/F814W & Dotter et al. (2011)       \\[+0.10cm]
Pyxis      & HST/ACS F606W/F814W & Dotter et al. (2011)       \\
           & CFHT/MegaCam g/r    & \citet{munozetal2018c}     \\[+0.10cm] 
Reticulum II & CFHT/MegaCam g/r & \citet{munozetal2018c}      \\[+0.10cm]
Rup 106    & HST/ACS F606W/F814W & Dotter et al. (2011)       \\[+0.10cm]
Segue 3    & CFHT/MegaCam g/r & \citet{munozetal2018c}        \\[+0.10cm]
Ter 7      & HST/ACS F606W/F814W & \citet{sarajedinietal2007} \\[+0.10cm]
Ter 8      & HST/ACS F606W/F814W & \citet{sarajedinietal2007} \\
           & Ground-based V/I    & \citet{stetsonetal2019}    \\[+0.10cm] 
Whiting 1  & Ground-based V/I    & \citet{valchevaetal2015}   \\[+0.10cm]
\hline
\end{tabular}
\end{table}

We take the input list of dwarf galaxy candidates from the recent overview article by \citet{simon2019}. To this list of 54 possible dwarf galaxies, we add Eridanus~III \citep{bechtoletal2015} and DES 1 \citep{luqueetal2016}. 
In addition, we investigate the star clusters Kim~1 \citep{kimetal2015}, Kim~2 \citep{kimjerjen2015b}, Koposov~1 and Koposov~2 \citep{koposovetal2007},
Mu\~{n}oz~1 \citep{munozetal2012} and Segue~3 \citep{belokurovetal2010}. We exclude from the list of \citet{simon2019} all systems with (stellar) relaxation times significantly larger than $10^{10}$ years since these
would be unsegregated even without dark matter. We also neglect systems more distant than about 120 kpc, since their main-sequence turn-offs are so faint that only a small part of their main sequence stars can be
observed, which does not allow us to detect mass segregation confidently. In addition we neglect systems that have absolute magnitudes less than about $M_V \approx -0.5$ since these objects have less than
50 observable stars on the main sequence, which again does not allow to detect mass segregation with high enough confidence. This leaves us with a list of 33 objects. Seven of these have published photometry and we list
their names as well as the sources for their photometry in Table~1.

For the remaining globular clusters and dwarf galaxies, we performed the photometric reduction ourselves. For most objects we used \hst images that we downloaded from the  {\tt STSci} archive. We performed stellar
photometry on these images using {\tt DOLPHOT} \citep{dolphin2000, dolphin2016}. For ACS and WFC3 observations, we performed the photometry on the CTE corrected flc images, while for the WFPC2 observations
we used the c0m images to perform the photometry. We used the point-spread functions provided for each camera and filter combination by {\tt DOLPHOT} for the photometric reductions.
After performing the photometry, we transformed the HST instrumental coordinates into equatorial ones by cross-matching stellar positions and magnitudes from HST with the positions of stars in the \gaia EDR3
catalogue \citep{brownetal2021}. We list the names of all globular clusters and ultra-faint dwarf galaxies that we analysed in Table~2 together with the \hst camera and filter combinations, the \hst
proposal ID and the total exposure times for each cluster. The exposure times are usually several thousand sec, which allows us to measure stellar magnitudes down to about 27th or 28th magnitude in
each passband.
\begin{table}
\caption{Observing log for the ground-based and \hst photometry analysed by us}
\footnotesize
\begin{tabular}{l@{$\;$}l@{$\!\!$}r@{}c}
\hline
Star Cluster/  & \multirow{2}{*}{Telescope/Camera/Filter} & \multicolumn{1}{c}{Proposal} & \multicolumn{1}{c}{Exp. time}\\
UFD &  & \multicolumn{1}{c}{ID} & (sec) \\
\hline
& & & \\[-0.20cm]
AM 1      & HST/WFPC2 F555W/F814W & $\;\,$6512 & 36240 \\[+0.10cm]
AM 4      & HST/WFC3 F606W/F814W  & 11680 & $\;\,$9840 \\[+0.10cm] 
Bo{\"o}tes II & HST/ACS F606W/F814W  & 14734 & 18388   \\[+0.10cm]
Cetus II      & HST/ACS F606W/F814W  & 14734 & $\;\,$9212 \\[+0.10cm]
Crater    & HST/ACS F606W/F814W & 13746 & $\;\,$8010      \\[+0.10cm]
DES 1         & Gemini/GMOS-S g,r & GS2016B & $\;\,$3600  \\[+0.10cm]
Draco II      & HST/ACS F606W/F814W  & 14734 & $\;\,$9324 \\[+0.10cm]
Eridanus  & HST/WFPC2 F555W/F814W & $\;\,$6106 & 17330    \\[+0.10cm]
Eridanus III  & HST/ACS F606W/F814W  & 14234 & 10340      \\[+0.10cm]
Grus I        & HST/ACS F606W/F814W  & 14734 & $\;\,$9532 \\[+0.10cm]
Horologium I  & HST/ACS F606W/F814W  & 14734 & $\;\,$9254 \\[+0.10cm]
Horologium II & HST/ACS F606W/F814W  & 14734 & 18508      \\[+0.10cm]
Laevens 3 & CFHT Megacam g/r & 15AD84 & $\;\,$4140        \\[+0.10cm]
NGC 2419  & HST/WFPC2 F606W/F814W & $\;\,$5481 & 20300 \\
          & HST/ACS F814W         & 10815 & $\;\,$4500 \\
          & HST/WFC3 F606W/F814W  & 11903 & $\;\,$2300 \\
          & HST/ACS F606W/F814W   & 14235 & 10000      \\[+0.10cm]
NGC 4147  & HST/ACS F606W/F814W   & 10775 & $\;\,$3500 \\[+0.10cm]
NGC 5053  & HST/WFPC2 F606W/F814W & 11125 & $\;\,$8800 \\
          & HST/WFC3 F606W/F814W  & 14235 & $\;\,$2120 \\[+0.10cm]
NGC 5139  & HST/ACS F435W/F625W   & $\;\,$9442 & $\;\,$6180 \\
          & HST/ACS F606W/F814W   & 10252 & $\;\,$3430 \\
          & HST/ACS F606W/F814W   & 12193 & $\;\,$1100 \\
          & HST/ACS F435W/F606W   & 13066 & $\;\,$3390 \\
          & HST/ACS F606W/F814W   & 15594 & $\;\,$6670 \\ 
          & HST/ACS F606W         & 15764 & $\;\,\;\,$680   \\[+0.10cm]
NGC 5466  & HST/WFPC2 F606W/F814W & 11125 & $\;\,$8800 \\
          & HST/WFC3 F606W/F814W  & 14235 & $\;\,$2080      \\[+0.10cm]
NGC 5897  & HST/WFPC2 F606W/F814W & 11125 & $\;\,$8800      \\[+0.10cm]
Pal 3     & HST/WFPC2 F555W/F814W & $\;\,$5672 & 14920      \\[+0.10cm]
Pal 4     & HST/WFPC2 F555W/F814W & $\;\,$5672 & $\;\,$8095 \\[+0.10cm]
Pal 13    & HST/WFC3 F606W/F814W  & 11680 & $\;\,$4900      \\[+0.10cm] 
Pal 14    & HST/WFPC2 F555W/F814W & $\;\,$6512 & $\;\,$9480 \\[+0.10cm] 
Phoenix II    & HST/ACS F606W/F814W  & 14734 & $\;\,$9254   \\[+0.10cm]
Pictoris I    & HST/ACS F606W/F814W  & 14734 & $\;\,$9532   \\[+0.10cm]
Reticulum III & HST/ACS F606W/F814W  & 14766 & 54492        \\[+0.10cm]
Segue 1       & HST/ACS F606W/F814W  & 14734 & 18420        \\[+0.10cm]
Segue 2       & HST/ACS F606W/F814W  & 14734 & $\;\,$9216   \\[+0.10cm]
Sgr II        & HST/ACS F606W/F814W  & 14734 & $\;\,$9216   \\[+0.10cm]
Triangulum II & HST/ACS F606W/F814W  & 14734 & 18384        \\[+0.10cm]
Tucana III    & HST/ACS F606W/F814W  & 14734 & 18584        \\[+0.10cm]
Tucana V      & HST/ACS F606W/F814W  & 14734 & $\;\,$9322   \\[+0.10cm]
Virgo I       & HST/ACS F606W/F814W  & 15332 & $\;\,$4752   \\[+0.10cm]
Willman 1     & HST/ACS F606W/F814W  & 14734 & $\;\,$9254   \\[+0.10cm]
\hline
\end{tabular}
\end{table}

For two objects, DES~1 and Laevens~3, we performed DAOPHOT PSF photometry on 
deep ground-based images. Laevens~3 was observed in the g and i bands for 
1440 and 2700 sec, respectively, with MegaCAM on CFHT. We used the advanced 
data products MegaPipe.612.210.G.MP9402 and MegaPipe.612.210.I.MP9703 from the
CFHT data archive for our photometry. The photometric calibration was performed
according to the prescription given in \citet{longeardetal2019}. DES~1 was 
observed in the g and r bands for 1800s each with GMOS-S on Gemini South. 
No advanced data products were available in the Gemini archive, so we reduced 
the raw data ourselves with the DRAGONS pipeline\footnote{\url{https://dragons.readthedocs.io/en/release-2.1.x/}}. The photometric calibration
was performed following the procedure described in \citet{connetal2018}.
The instrumental coordinates of both objects were transformed to the Gaia 
system by cross-matching bright stars with the Gaia EDR3 catalogue \citep{brownetal2021}.

For each analysed cluster, we also estimated the photometric completeness using artificial star tests. For the clusters with HST data, we created 
75,000 artificial stars in each cluster, equally distributed across
the HST field. The artificial stars were also equally distributed in magnitude along the location of the cluster main sequence from the turnover down to the faintest detectable magnitudes.
We then used the {\tt DOLPHOT} {\tt fakestars} task to recover the 
magnitudes of the artificial stars and applied the same quality cuts to the artificial stars that we used to select bona-fide stars in the observed data set. Stars were counted as recovered if 
their derived magnitudes were within 0.2 mag of the input ones in both magnitude bands. We then estimated the completeness fraction for each observed star from the ratio of successfully recovered stars to 
all inserted stars using the nearest 20 artificial stars that are within 0.2 mag of the magnitude of each observed star.  

The incompleteness tests for the ground-based data were performed with the DAOPHOT package artdata (tasks starlist and addstars). 100 artificial stars with random positions and random magnitudes in the range of the main sequence were added to the images of the individual filters, and photometry 
was performed in the same manner as for the true stars. The same photometric selection criteria were applied to define the final sample of valid stars. 
This experiment was repeated 1000 times per image and filter so that the final incompleteness values are based on the statistics of $10^5$ artificial stars.

\begin{figure}
\begin{center}
\includegraphics[width=0.97\columnwidth]{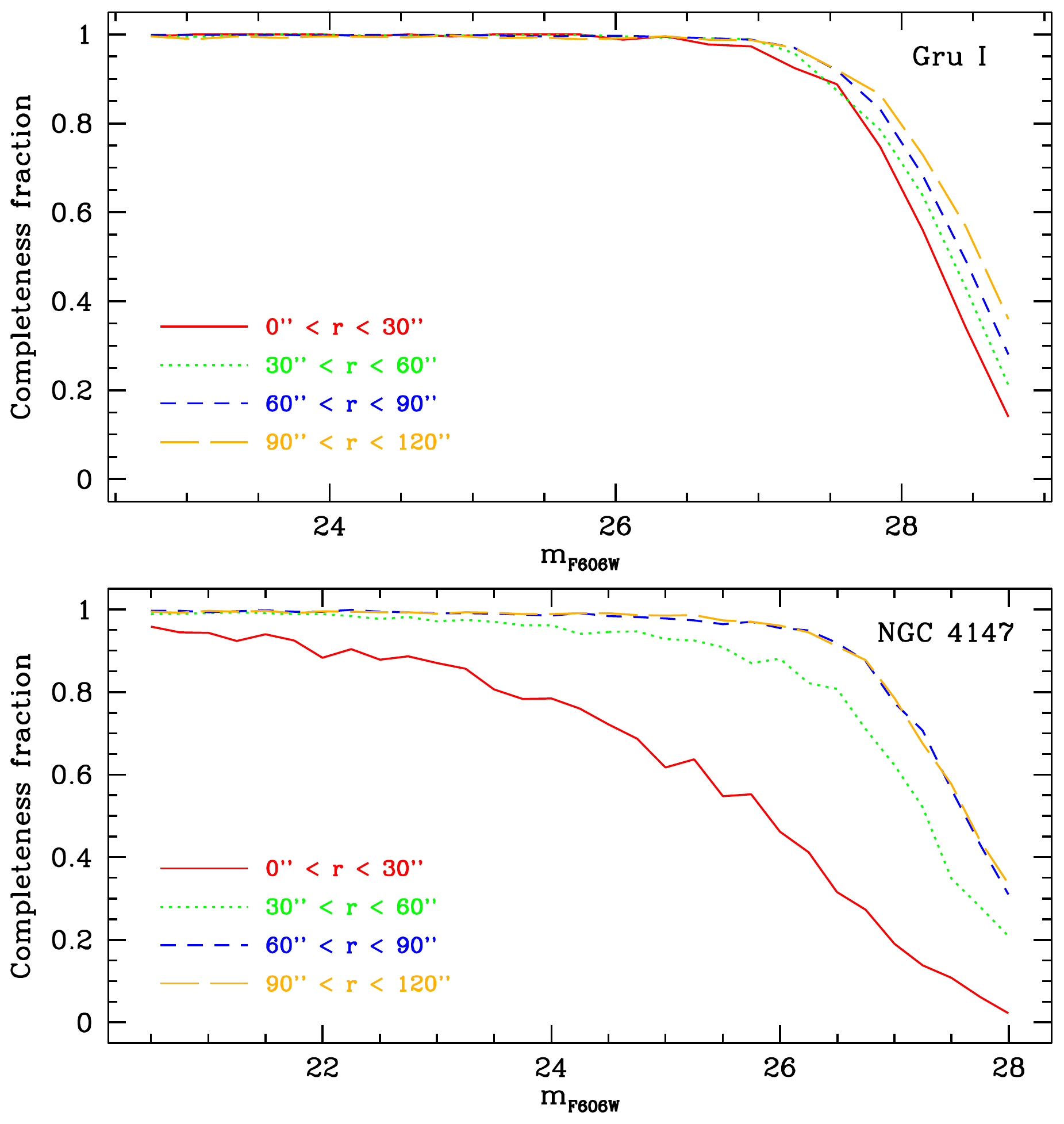}
\caption{Photometric completeness as a function of F606W magnitude for four different distance ranges from the centre for the ultrafaint dwarf galaxy candidate Grus~I (top panel) and the globular cluster NGC~4147 (bottom panel). Due to the low stellar
 density of Grus~I, the photometric completeness is nearly independent of distance, while the higher stellar density in the centre of NGC~4147 decreases the completeness at the faint end.}
\label{fig:compl}
\end{center}
\end{figure}

Fig.~\ref{fig:compl} shows the results of our completeness tests 
for two representative cases, Grus~I and NGC~4147. It can be seen that for Grus~I our observations are more than 90\% complete down to F606W=27.5 mag, well below the limit that we later 
use to determine mass segregation. Due to the low stellar density of Grus~I, the completeness fraction is also nearly independent of the distance to the centre. We obtain similar results 
for the other dwarf galaxies and also most globular clusters. In NGC~4147 our observations are also more than 90\% complete down to F606W=26.5, the limit of our later analysis, for radii $r>$30''. 
The completeness drops below 90\% only for the faintest stars and only for radii $r<$30''. We estimated the photometric completeness for all clusters for which we performed the photometry ourselves. 
For the other globular clusters we used the completeness data from \citet{sarajedinietal2007} for the inner fields.
When calculating cumulative profiles, we use the inverse of the completeness fraction as a weighting factor to correct for the
incompleteness, as will be explained below.

\section{Mass segregation determination}

\begin{figure*}[h]
\begin{center}
\includegraphics[width=0.95\textwidth]{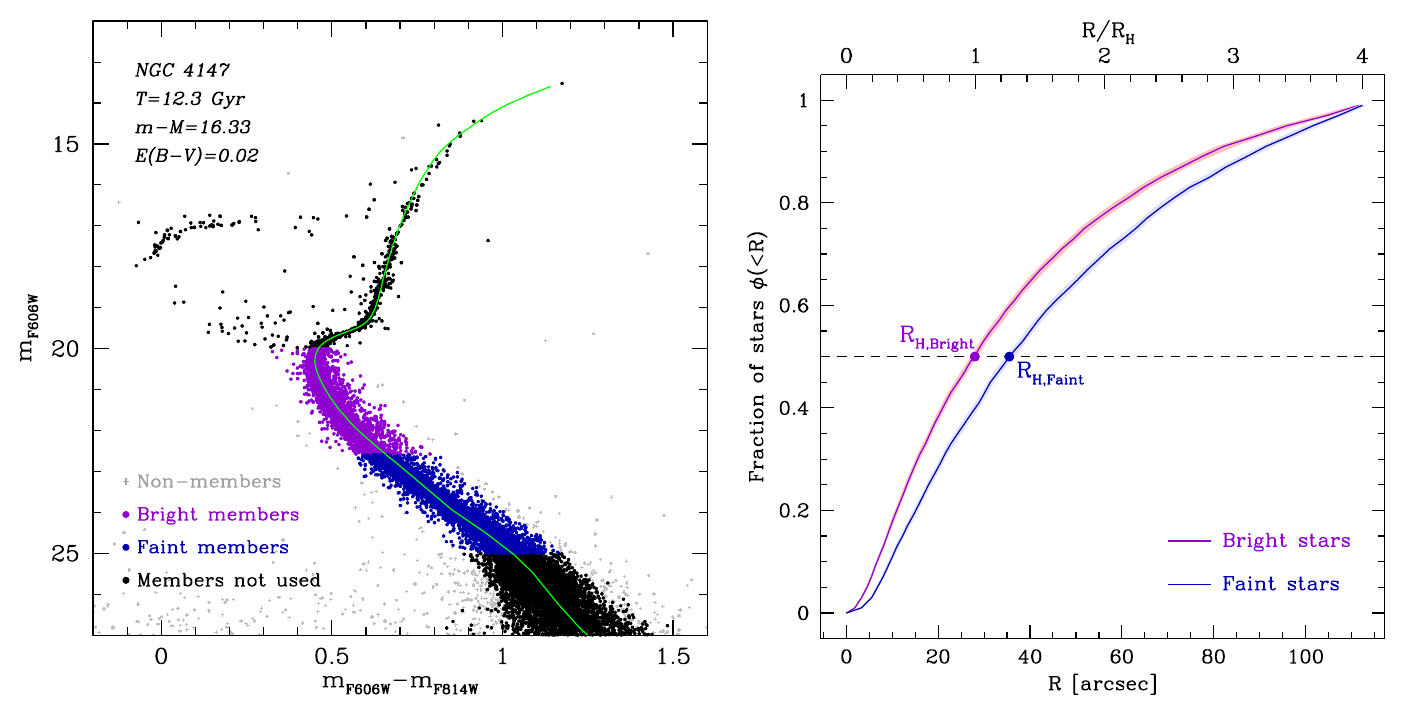}
\caption{Colour magnitude diagram depicting our selection of cluster members (left panel) and the resulting cumulative distribution of stars (right panel) for NGC 4147. The green line shows the PARSEC isochrone that
we used to  select the cluster members. The parameters of this isochrone are given in the upper left corner of the left panel. Red and blue stars show the cluster stars selected to calculate the mass segregation
profile. The right panel shows the cumulative profile after correcting for incompleteness of stars at the faint end. The solid lines depict the cumulative distribution that we derive from the data, while the shaded
areas show the region containing two thirds of all curves when doing the bootstrapping.}
\label{fig:mseg}
\end{center}
\end{figure*}

In order to determine the amount of mass segregation for each cluster/dwarf galaxy, we first fitted the cluster colour-magnitude diagrams (CMDs) 
with PARSEC isochrones \citep{bressanetal2012}. We took the metallicity and reddening of each globular cluster from the 2010 edition of \citet{harris1996}, while
for star clusters and UFDs we took the metallicity from published literature, or, if no metallicity determination was available, we assumed a metallicity of [Fe/H]=\mbox{-2.0.} 
We assumed cluster ages of 5 Gyr for Whiting~1 \citep{carraroetal2007}, 8.0 Gyr for Ter~7 \citep{dotteretal2010} and 8.0 Gyr for Pal~1 \citep{rosenbergetal1998}. For all other globular clusters we assumed
an age of 12 Gyr, 
which is compatible with the published ages of these systems \citep[e.g][]{dotteretal2010,vandenbergetal2013}. We then created PARSEC isochrones for each cluster for the same camera and filter combination for which
we have observed data.  The isochrones were then shifted in distance modulus and reddening until we obtained the best match with the observed color-magnitude diagram of each cluster. We then used 
the isochrone to select the cluster members. Cluster members were required to deviate by no more than twice their photometric error or 0.10 mag, whichever was larger, from the isochrone. 
Fig.~\ref{fig:mseg} depicts the selection of cluster members and non-members for NGC 4147.

After selecting the cluster members, we determined for each star a weight factor $w_i$ according to:
\begin{equation}
 w_i = \frac{1}{f_{r,i}} \frac{1}{f_{p,i}}
\label{eq:weight}
\end{equation}
where $f_{r,i}$ is the fraction of the cluster that is covered by our photometry at the distance of star $i$ and $f_{p,i}$ is the fraction of artificial stars recovered in our artificial star tests
with a magnitude and color similar to star $i$ and at a similar spatial position. We determined the fraction covered by the photometry in steps of 1 arcsec from the cluster centre by distributing
360 points uniformly along a circle around the cluster centre. For each of these points we then determined the distance of the nearest star in our photometry from this point. If this distance was
less than a critical distance (usually 1 arcsec but we chose slightly higher values for clusters with fewer member stars), we considered the point to be covered by our photometry. The spatial completeness factor 
$f_{r,i}$ for each radius is then given by the fraction of the points that were found to be covered. We stopped the analysis at a maximum radius $R_{max}$ where either the density of cluster stars becomes too low
or the photometric completeness $f_{r,i}$ drops below 15\%. The latter was done in order to avoid giving individual stars too much weight in the determination of the mass segregation profile.
The values of $R_{max}$ are given in Table 3 for each system. For most systems they are at least two times the projected half-light radius of the system.
The photometric completeness factor $f_{p,i}$ was determined from the 20 nearest artificial stars that were within $\pm 0.20$ mag of the short-wavelength (usually F606W) magnitude of each star.

After the determination of the weight factors, we ordered the member stars according to their brightness and splitted the sample
into two equally large groups according to stellar brightness. We then ordered the stars of each group according to their distance from the centre and determined their cumulative distribution according to
\begin{equation}
  \phi(<R) = \frac{\sum_i^{N(<R)} w_i}{\sum_i^{N_{Tot}} w_i}
\end{equation}
where the sum in the numerator runs over all bright/faint stars up to a given radius $R$ and the sum in the denominator runs over all stars of each group. 
For both the faint and bright sample, we then calculated
the radius $R_H$ where $\phi(<R)$ is equal to 0.5. We then calculated the ratio of these radii $r=R_{H,Bright}/R_{H,Faint}$ and determined its error by boot-strapping. Fig.~\ref{fig:mseg} shows the
result of our mass segregation analysis for NGC~4147, for which we find $R_{H,Bright}=30.4 \pm 0.55$ arcsec, $R_{H,Faint} = 41.9 \pm 0.64$ arcsec and a ratio of $R_{H,Bright}/R_{H,Faint}=0.73 \pm 0.02$, i.e. a highly significant detection of mass segregation.

As an additional measure of mass segregation, we perform a KS test on the cumulative distribution of bright and faint stars to test the significance that both distributions are different from each other, i.e. that the
system is mass segregated. The corresponding probabilities are given in column~12 of Table~3. Since the stars carry individual weight factors, we modify the standard two-sample KS test as described by \citet{monahan2001}
sec.~12.4 by replacing the sample sizes $n_1$ and $n_2$ in the effective sample size factor
\begin{equation}
 n = \sqrt{\frac{n_1 n_2}{n_1+n_2}}
\end{equation}
by
\begin{equation}
 n_1 = \frac{\left(\sum_i w_{b,i} \right)^2}{\sum_i w_{b,i}^2}
\end{equation}
and
\begin{equation}
 n_2 = \frac{\left(\sum_i w_{f,i} \right)^2}{\sum_i w_{f,i}^2}
\end{equation}
where $w_{b,i}$ and $w_{f,i}$ are the weight factors from eq.~\ref{eq:weight} for the bright and faint stars respectively.

We also determine the relaxation time of each globular cluster and dwarf galaxy candidate. For the globular clusters we take the relaxation times from \citet{baumgardtetal2019a}, updated for the new distances 
derived by \citet{baumgardtvasiliev2021}.  \citet{baumgardtetal2019a} fitted a large set of $N$-body models to the
surface density, velocity dispersion profiles and the stellar mass functions of globular clusters, thereby creating a detailed model for each individual globular cluster. For star clusters and dwarf galaxies not studied by 
\citet{baumgardtetal2019a}, we estimate the relaxation times according to \citep{spitzer1987}:
\begin{equation}
T_{RH} = 0.138 \frac{\sqrt{M} r_h^{1.5}}{\sqrt{G} \langle m \rangle \ln{\gamma N}}
\end{equation}
where $M$ is the (stellar) mass of the system, $r_h$ the 3-dimensional half-mass radius, $\langle m \rangle$ the average mass of stars and $N=M/\langle m \rangle$ the number of stars. $\gamma$ is a constant in the Coulomb logarithm for which we assume $\gamma=0.11$
\citep{gierszheggie1994}. We calculate the mass of the star clusters/dwarf galaxies from their V-band luminosities, assuming a mass-to-light ratio of $M/L_V = 1.8 M_\odot/L_\odot$ \citep{baumgardt2017}. We also assume $\langle m \rangle=0.35$~$M_\odot$
and $r_h=0.75 r_{hp}$ for the conversion of projected half-light radii to 3D ones.
The above formula is valid for stellar systems free of dark matter. In the presence of dark matter the relaxation times would be significantly larger since ultra-faint dwarf galaxies are expected to be strongly dark matter
dominated \citep{simongeha2007}. Approximating the relaxation process as dynamical friction of stars against the much lighter dark matter particles and using eq.~7-26 of \citet{binneytremaine1987}, we find that the relaxation time should increase approximately with
the square root of the ratio of dark to luminous matter $T_{RH} \sim \sqrt{M_{DM}/M_{*}}$, and would therefore be a factor 30 larger for a system that has 1000 times as much dark matter than luminous matter. With the exception of the 
smallest systems that have relaxation times less than about 0.3 Gyr, we would therefore not expect to detect mass segregation in any system that is dark matter dominated. 

\section{$N$-body simulations of mass segregation}

In order to compare the observed amount of mass segregation with theoretical expectations, we analysed the amount of mass segregation in the $N$-body simulations done by \citet{baumgardtsollima2017} and \citet{baumgardtmakino2003}.
\citet{baumgardtsollima2017} performed a set of 16 $N$-body simulations of star clusters evolving in the tidal field of the Milky Way. The simulated clusters contained between $65,536$ and $200,000$
stars that followed a \citet{kroupa2001} mass function initially. In the simulations, the initial cluster orbits, cluster sizes and the fraction of the assumed black hole retention fraction were varied.
While most simulations assumed 10\% retention fraction for the black holes, five runs had higher black hole retention fractions up to 100\%.
The simulations by \citet{baumgardtmakino2003} were similar in set-up and we analysed all simulations that started with $N =131,072$ stars and survived for a Hubble time from their paper.
In order to also cover star clusters resembling extended halo globular clusters, we performed an additional three simulations with a set-up similar to run 4 of \citet{baumgardtsollima2017}
(Kroupa mass function, $N=131,072$ stars, 10\% black hole retention fraction) but orbiting at a galactocentric distance of $R_G=100$ kpc and starting with initial half-mass radii of
10~pc, 15~pc and 25~pc respectively. At an age of 12 Gyr the simulated clusters had typical half-mass radii of 5 to 35 pc and a remaining bound mass of between 5,000 and 30,000 $M_\odot$. They are therefore
comparable in size to the lower-mass star clusters analysed here. In addition, their mass functions have flattened due to mass segregation and preferential mass loss of low-mass stars. When fitted by a power-law mass function,
we find slopes between $\alpha = -0.80$ and $-1.30$ for most of them, in good agreement with what has been found for mass functions of globular clusters \citep[e.g][]{baumgardtsollima2017} and ultra-faint dwarf galaxies \citep[e.g][]{gehaetal2013,gennaroetal2018}.

For each cluster in the $N$-body simulations, we analysed only the snapshots between 10 and 13 Gyrs. For each of these $\sim$1,500 snapshots, we splitted the main sequence stars above a certain mass-cutoff $M_{Low}$
into bright and faint stars and determined the $R_{H,Bright}/R_{H,Faint}$ ratio for each individual snapshot in the same way as we did for the observed globular clusters. We also
determined the relaxation time and the $T_{Age}/T_{RH}$ ratio of each snapshot. Since the amount of mass segregation and the mass segregation ratio will depend on the lower mass limit, 
we repeat the analysis for four different lower mass limits of $M_{Low,NBODY} = 0.35$ M$_\odot$, 0.45 M$_\odot$, 0.55 M$_\odot$ and 0.65 M$_\odot$ respectively.

Fig.~3 depicts the resulting mass segregation ratios for the four different cases. The black lines and grey shaded regions in Fig. 3 show the average ratio of $R_{H,Bright}/R_{H,Faint}$ and the region containing 90\% of all snapshots for the $N$-body simulations. Systems that have relaxation times much longer than their ages 
have insufficient time to become mass segregated
and therefore still have mass segregation ratios near unity. Significant mass segregation only sets in around $T_{Age}/T_{RH}=3$ and is largely complete around $T_{Age}/T_{RH}=10$, at which point $R_{H,Bright}/R_{H,Faint}$
levels off at values between 0.70 to 0.85 depending on the mass range of stars that are considered. The data shown in Fig.~3 is for the case that the stellar distribution is analysed out to the tidal radius of each system.  Decreasing the value of $R_{Max}$, we find that the resulting mass segregation ratio will be closer to unity. If $R_{max}$ is only twice the projected half-light radius $r_{h,p}$ for example we find that $R_{H,Bright}/R_{H,Faint}$ 
increases by about 0.05
for the most mass segregated clusters with $T_{Age}/T_{RH}>10$ and by about 0.10 if the stellar distribution is analysed only up to $R_{max}=r_{h,p}$. Changes are 
 similar but smaller for smaller values of $T_{Age}/T_{RH}$. 

$N$-body simulations show that star clusters that start with primordial mass segregation need several relaxation times to erase the mass segregation via two-body relaxation
\citep{pavliketal2019,pavlik2020}. Our results therefore strongly argue against primordial mass segregation among low mass stars in globular clusters.
\begin{figure*}
\begin{center}
\includegraphics[width=0.95\textwidth]{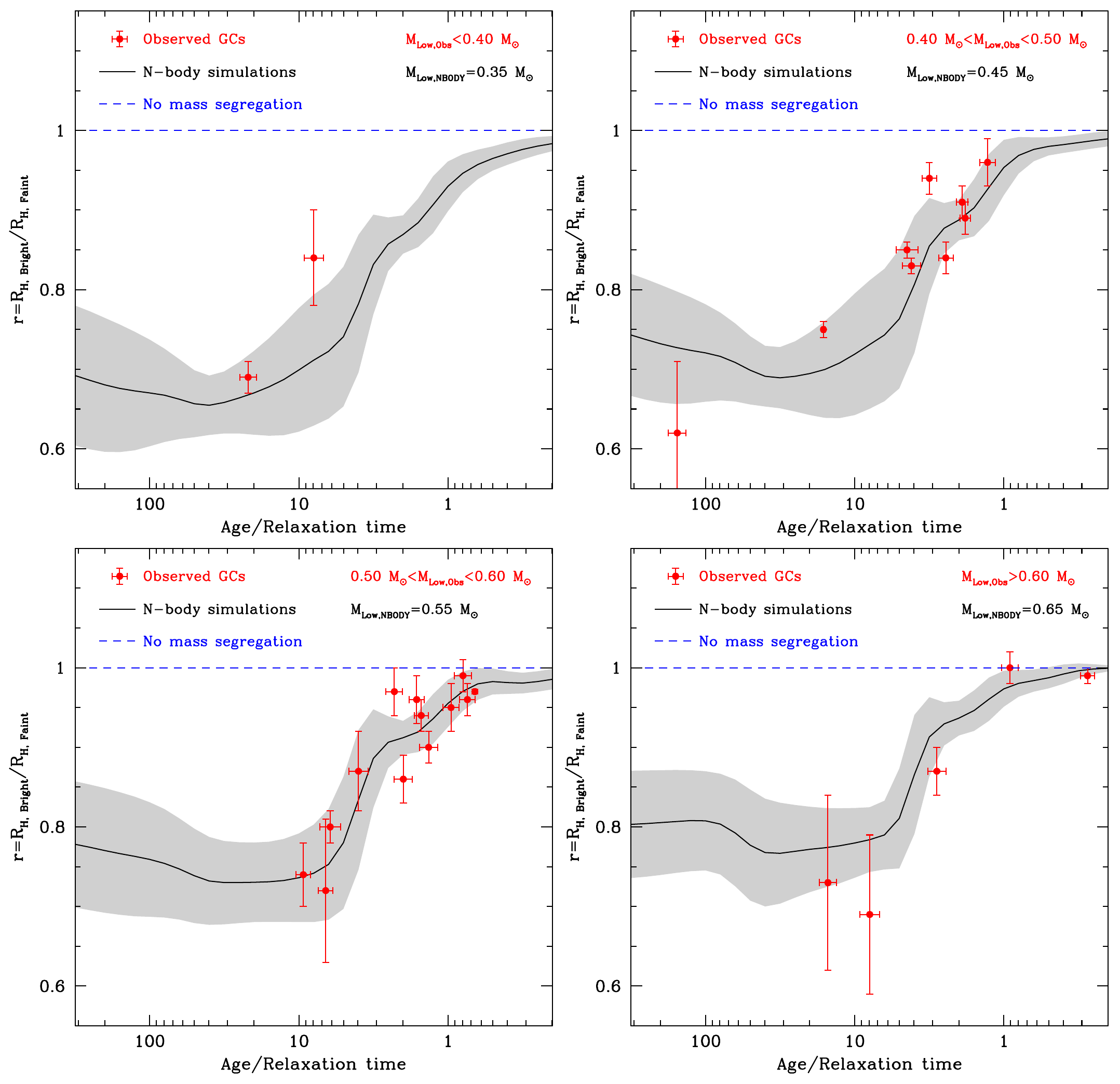}
\caption{Ratio of the radius containing half the bright stars, $R_{H,Bright}$, over the radius containing half the faint stars, $R_{H,Faint}$, as a function of the dynamical age $T_{Age}/T_{RH}$ for Milky Way globular clusters.
 The globular sample is split into four different groups depending on the mass $M_{Low,Obs}$ of the lowest-mass star that was analysed in each cluster. The lowest mass star analysed in the N-body simulations $M_{Low,NBODY}$ is varied
  accordingly and is indicated in each panel.
 Observed globular clusters are shown by red circles with error bars, while the results of the $N$-body simulations are shown as a black line and grey
shaded region. Observed globular clusters with young dynamical ages $T_{Age}/T_{RH} \le 1$ are almost completely unsegregated with $R_{H,Bright}$ only a few percent smaller than $R_{H,Faint}$.
The amount of mass segregation increases with increasing dynamical age and the range of values seen for each $T_{Age}/T_{RH}$ is in good agreement with $N$-body simulations of initially unsegregated clusters,
 showing that the observed mass segregation is not primordial but a result of two-body relaxation driven dynamical evolution.}
\label{fig:mseggc}
\end{center}
\end{figure*}

\section{Results}
\label{sec:results}

\begin{table*}
\caption{Adopted parameters, relaxation times and mass segregation results for the studied clusters and ultra-faint dwarf galaxies}
\centering
\resizebox{\textwidth}{!}{\begin{tabular}{lcccccccccrccc}
\hline
 \multirow{2}{*}{Cluster} &  \multirow{2}{*}{$M_V$} & $R_H$ & $D_\odot$ & Mass       & $T_{RH}$ & $m_{Low}$ & $m_{Med}$ & $m_{Up}$ & $R_{max}$ & \multirow{2}{*}{$N_{Star}$} & $P_{Mseg}$ &  \multirow{2}{*}{$R_{Bright}/R_{Faint}$} &  \multirow{2}{*}{Ref.}\\
          &       & ['']  &  [kpc]    & [$M_\odot$] & [Gyr] & [$M_\odot$] & [$M_\odot$] & [$M_\odot$] & [''] & & [\%] & & \\     
 & & \\[-0.2cm]
\hline
 & & \\[-0.3cm]
\multicolumn{14}{c}{Globular clusters} \\
 & & \\[-0.3cm]
\hline
AM 1      & -6.11 & $\;\;$25.8 &      118.9 & $4.0 \cdot 10^4$ & $\;\,$4.68 & 0.64 & 0.74 & 0.80 & $\;\,$82 &  2190 & 100.0 & $0.87 \pm 0.03$ & 1,2,3 \\  
AM 4      & -0.97 & $\;\;$45.4 & $\;\;$29.0 & $7.5 \cdot 10^2$ & $\;\,$1.51 & 0.27 & 0.50 & 0.88 & $\;\,$97 &   558 & $\;\;$98.6 & $0.84 \pm 0.06$ & 1,2,3 \\ 
Crater    & -5.10 & $\;\;$27.1 &      147.2 & $1.1 \cdot 10^4$ & $\;\,$5.75 & 0.60 & 0.75 & 0.92 & 115 &  2976 &  100.0 & $0.86 \pm 0.03$ & 1,2,3 \\
Eridanus  & -5.43 & $\;\;$33.5 & $\;\;$84.7 & $1.2 \cdot 10^4$ & $\;\,$2.75 & 0.54 & 0.69 & 0.81 &  95 &  1513 &  100.0 & $0.87 \pm 0.05$ & 1,2,3 \\
Laevens 3 & -2.85 & $\;\;$23.8 & $\;\;$61.8 & $2.1 \cdot 10^3$ & $\;\,$0.78 & 0.70 & 0.74 & 0.78 &  70 &   197 &  $\;\;$96.6 & $0.74 \pm 0.12$ & 1,3,4 \\
NGC 1261  & -7.80 & $\;\;$40.9 & $\;\;$16.4 & $1.8 \cdot 10^5$ & $\;\,$1.91 & 0.49 & 0.65 & 0.80 & 130 & 44171 &  100.0 & $0.77 \pm 0.01$ & 1,2,3 \\
NGC 2419  & -9.17 & $\;\;$45.8 & $\;\;$88.5 & $9.8 \cdot 10^5$ &      48.98 & 0.67 & 0.74 & 0.80 & 160 & 51855 &  $\;\;$99.0 & $0.99 \pm 0.01$ & 1,2,3 \\
NGC 4147  & -6.11 & $\;\;$28.1 & $\;\;$18.5 & $3.8 \cdot 10^4$ & $\;\,$0.37 & 0.40 & 0.61 & 0.78 & 139 & 12214 &  100.0 & $0.69 \pm 0.02$ & 1,2,3 \\
NGC 5053  & -6.33 &      145.5 & $\;\;$17.5 & $7.4 \cdot 10^4$ & $\;\,$9.12 & 0.48 & 0.68 & 0.78 & 442 &  8327 &  $\;\;$99.9 & $0.96 \pm 0.03$ & 1,2,3 \\
NGC 5139  &$\!\!$-10.17 &286.8 & $\;\;$5.43 & $3.9 \cdot 10^6$ &      26.91 & 0.52 & 0.64 & 0.78 & 474 &327420 &  100.0 & $0.97 \pm 0.00$ & 1,2,3 \\ 
NGC 5466  & -6.72 &      122.1 & $\;\;$16.1 & $6.0 \cdot 10^4$ & $\;\,$6.46 & 0.46 & 0.60 & 0.79 & 434 & 12055 &  $\;\;$98.4 & $0.91 \pm 0.02$ & 1,2,3 \\
NGC 5897  & -7.29 &      124.6 & $\;\;$12.6 & $1.5 \cdot 10^5$ & $\;\,$5.37 & 0.44 & 0.60 & 0.79 & 408 & 27076 &  100.0 & $0.89 \pm 0.02$ & 1,2,3 \\
NGC 6121  & -6.77 &      277.6 & $\;\;$1.85 & $8.8 \cdot 10^4$ & $\;\,$0.78 & 0.46 & 0.66 & 0.82 & 600 & 23449 &  100.0 & $0.75 \pm 0.01$ & 1,2,3 \\
NGC 6426  & -6.57 & $\;\;$51.9 & $\;\;$20.7 & $7.0 \cdot 10^4$ & $\;\,$2.45 & 0.41 & 0.56 & 0.79 & 130 & 29678 &  100.0 & $0.85 \pm 0.01$ & 1,2,3 \\
NGC 7006  & -7.50 & $\;\;$22.6 & $\;\;$39.3 & $1.3 \cdot 10^5$ & $\;\,$2.34 & 0.49 & 0.65 & 0.80 & 113 & 41596 &  100.0 & $0.83 \pm 0.01$ & 1,2,3 \\
NGC 7492  & -5.80 & $\;\;$63.9 & $\;\;$24.4 & $2.7 \cdot 10^4$ & $\;\,$1.95 & 0.55 & 0.69 & 0.80 & 200 &  4544 &  100.0 & $0.80 \pm 0.02$ & 1,2,3 \\
Pal 1     & -1.76 & $\;\;$33.2 & $\;\;$11.2 & $1.0 \cdot 10^3$ & $\;\,$0.08 & 0.42 & 0.62 & 0.91 & 124 &   446 &  100.0 & $0.62 \pm 0.09$ & 1,2,3 \\
Pal 3     & -5.49 & $\;\;$44.0 & $\;\;$94.8 & $1.4 \cdot 10^4$ & $\;\,$7.94 & 0.54 & 0.66 & 0.79 & $\;\,$79 &  3547 & $\;\;$99.6 & $0.90 \pm 0.02$ & 1,2,3 \\
Pal 4     & -5.86 & $\;\;$32.2 &      101.4 & $2.6 \cdot 10^4$ & $\;\,$8.71 & 0.58 & 0.70 & 0.82 & 136 &  2491 & $\;\;$79.9 & $0.97 \pm 0.03$ & 1,2,3 \\
Pal 5     & -4.94 & $\;\;$21.9 & $\;\;$21.9 & $1.0 \cdot 10^4$ & $\;\,$7.41 & 0.58 & 0.70 & 0.80 & 300 &  2221 & $\;\;$97.2 & $0.96 \pm 0.03$ & 1,2,3 \\ 
Pal 12    & -4.41 & $\;\;$18.5 & $\;\;$76.7 & $6.3 \cdot 10^3$ & $\;\,$1.29 & 0.54 & 0.69 & 0.88 & 300 &  2020 & 100.0 & $0.74 \pm 0.04$ & 1,2,3 \\
Pal 13    & -3.15 &      113.0 & $\;\;$23.5 & $3.0 \cdot 10^3$ & $\;\,$1.82 & 0.51 & 0.66 & 0.77 & 250 &   584 & 100.0 & $0.72 \pm 0.09$ & 1,2,3 \\
Pal 14    & -5.30 & $\;\;$77.0 & $\;\;$73.6 & $1.9 \cdot 10^4$ &      13.80 & 0.53 & 0.68 & 0.82 & 112 &  2217 & $\;\;$68.6 & $0.95 \pm 0.03$ & 1,2,3 \\
Pal 15    & -5.72 & $\;\;$94.0 & $\;\;$44.1 & $5.0 \cdot 10^4$ &      12.02 & 0.55 & 0.68 & 0.79 & 134 &  6634 & $\;\;$37.8 & $0.99 \pm 0.02$ & 1,2,3 \\
Pyxis     & -5.49 & $\;\;$96.2 & $\;\;$36.5 & $2.5 \cdot 10^4$ & $\;\,$7.76 & 0.60 & 0.71 & 0.82 & 250 &  4566 & 100.0 & $0.94 \pm 0.02$ & 1,2,3 \\
Rup 106   & -6.15 & $\;\;$75.7 & $\;\;$20.7 & $4.2 \cdot 10^4$ & $\;\,$2.67 & 0.47 & 0.63 & 0.81 & 133 &  9111 & 100.0 & $0.94 \pm 0.02$ & 1,2,3 \\
Ter 7     & -5.28 & $\;\;$54.5 & $\;\;$24.3 & $2.1 \cdot 10^4$ & $\;\,$3.72 & 0.49 & 0.65 & 0.94 & 128 &  6752 & 100.0 & $0.85 \pm 0.02$ & 1,2,3 \\
Ter 8     & -6.52 & $\;\;$27.5 &      113.7 & $6.1 \cdot 10^4$ &      10.72 & 0.62 & 0.70 & 0.79 & 220 &  7434 & $\;\;$78.4 & $1.00 \pm 0.02$ & 1,2,3 \\
Whiting 1 & -4.17 & $\;\;$64.1 & $\;\;$30.6 & $2.0 \cdot 10^3$ & $\;\,$1.51 & 0.66 & 0.85 & 1.04 & 200 &   333 & 100.0 & $0.69 \pm 0.10$ & 1,2,3 \\
\hline
 & & \\[-0.3cm]
\multicolumn{14}{c}{Ultra-faint dwarf galaxies/Star clusters} \\
 & & \\[-0.3cm]
\hline
Balbinot 1      & -1.21 & $\;\;$52.2 & $\;\;$31.9 & $4.7 \cdot 10^2$  & $\;\,$0.82 & 0.55 & 0.63 & 0.78 &  80 & 141 &  $\;\;$99.8 & $0.65 \pm 0.10$ & 5,6 \\
Bo{\"o}tes II   & -2.94 &      190.2 & $\;\;$42.0 & $1.4 \cdot 10^3$  & $\;\,$5.63 & 0.41 & 0.54 & 0.78 & 225 & 285 &  $\;\;$54.0 & $0.95 \pm 0.07$ & 5,7 \\ 
Cetus II        & $\;\;$0.00 & 114.0 & $\;\;$26.3 & $1.5 \cdot 10^2$  & $\;\,$1.47 & 0.31 & 0.49 & 0.79 & 140 & 111 &  $\;\;$26.0 & $1.12 \pm 0.20$ & 5,8 \\
DES 1           & -1.42 & $\;\;$14.7 & $\;\;$76.0 & $5.7 \cdot 10^2$  & $\;\,$0.47 & 0.68 & 0.74 & 0.78 &  40 &  56 &  $\;\;$95.3 & $0.69 \pm 0.21$ & 9  \\
Draco II        & -2.90 &      162.0 & $\;\;$20.0 & $2.2 \cdot 10^3$  & $\;\,$3.17 & 0.25 & 0.43 & 0.78 & 170 & 150 &  $\;\;$88.4 & $0.78 \pm 0.07$ & 10 \\
Eridanus III    & -2.37 & $\;\;$18.0 & $\;\;$87.0 & $1.4 \cdot 10^3$  & $\;\,$1.05 & 0.53 & 0.65 & 0.78 & 130 & 440 &  $\;\;$91.7 & $0.84 \pm 0.09$ & 5 \\
Grus I          & -3.40 &      106.6 &      120.0 & $3.5 \cdot 10^3$  &      26.92 & 0.58 & 0.67 & 0.79 & 130 & 531 &  $\;\;$75.7 & $0.89 \pm 0.09$ & 11 \\
Grus II         & -3.90 &      360.0 & $\;\;$55.0 & $5.6 \cdot 10^3$  &      74.35 & 0.42 & 0.56 & 0.79 & 135 & 291 &   $\;\;\;\;$8.9 & $1.00 \pm 0.07$ & 12,13\\
Horologium I    & -3.40 & $\;\;$69.8 & $\;\;$68.0 & $3.5 \cdot 10^3$  & $\;\,$7.60 & 0.50 & 0.62 & 0.79 & 140 &1099 &   $\;\;\;\;$9.4 & $0.99 \pm 0.06$ & 14 \\
Horologium II   & -1.56 &      116.4 & $\;\;$78.0 & $6.5 \cdot 10^2$  & $\;\,$4.84 & 0.54 & 0.64 & 0.79 & 220 & 352 &  $\;\;$29.1 & $1.06 \pm 0.12$ & 5,15 \\
Kim 2           & -1.50 & $\;\;$28.8 &      100.0 & $6.1 \cdot 10^2$  & $\;\,$1.77 & 0.63 & 0.73 & 0.79 &  75 &  65 &  $\;\;$70.3 & $0.81 \pm 0.14$ & 5,16 \\
Koposov 1       & -1.04 & $\;\;$37.2 & $\;\;$48.3 & $4.0 \cdot 10^2$  & $\;\,$0.88 & 0.55 & 0.70 & 0.78 & 100 &  60 &  $\;\;$74.4 & $0.70 \pm 0.16$ & 5 \\
Koposov 2       & -0.92 & $\;\;$26.4 & $\;\;$34.7 & $3.6 \cdot 10^2$  & $\;\,$0.31 & 0.48 & 0.60 & 0.79 & 130 & 114 &  $\;\;$89.5 & $0.73 \pm 0.17$ & 5 \\
Pegasus III     & -2.95 & $\;\;$51.0 &      174.0 & $2.3 \cdot 10^3$  &      16.86 & 0.69 & 0.74 & 0.79 & 145 & 605 &  $\;\;$44.3 & $1.04 \pm 0.06$ & 17,18 \\
Phoenix II      & -2.80 & $\;\;$65.4 &       83.0 & $2.0 \cdot 10^3$  & $\;\,$7.69 & 0.55 & 0.64 & 0.78 & 135 & 491 &  $\;\;$19.5 & $1.07 \pm 0.07$ & 5,19 \\
Pictoris I      & -3.45 & $\;\;$52.8 &      110.0 & $3.7 \cdot 10^3$  &      10.52 & 0.63 & 0.70 & 0.78 & 125 & 585 &   $\;\;\;\;$7.4 & $1.02 \pm 0.08$ & 5,13 \\
Reticulum II    & -3.88 &      331.2 & $\;\;$31.4 & $5.1 \cdot 10^3$  &      28.23 & 0.44 & 0.57 & 0.78 & 550 &1953 &  $\;\;$25.6 & $1.03 \pm 0.05$ & 5,20 \\
Reticulum III   & -3.30 &      159.0 & $\;\;$92.0 & $3.2 \cdot 10^3$  &      34.27 & 0.55 & 0.66 & 0.79 & 130 & 450 &  $\;\;$92.5 & $1.20 \pm 0.10$ & 11 \\
Segue 1         & -1.30 &      217.2 & $\;\;$23.0 & $5.1 \cdot 10^2$  & $\;\,$4.84 & 0.27 & 0.47 & 0.78 & 240 & 292 &  $\;\;$40.6 & $0.93 \pm 0.06$ & 5 \\
Segue 2         & -1.86 &      225.6 & $\;\;$35.0 & $8.5 \cdot 10^2$  & $\;\,$9.65 & 0.32 & 0.49 & 0.78 & 155 & 336 &   $\;\;\;\;$4.5 & $1.02 \pm 0.06$ & 5 \\
Segue 3         & -0.87 & $\;\;$29.4 & $\;\;$27.0 & $3.4 \cdot 10^2$  & $\;\,$0.24 & 0.42 & 0.61 & 0.79 & 100 & 100 &  $\;\;$75.9 & $0.73 \pm 0.17$ & 5 \\
Sagittarius II  & -5.41 & $\;\;$89.9 & $\;\;$66.4 & $2.3 \cdot 10^4$  & $\;\,$8.37 & 0.57 & 0.67 & 0.78 & 200 &5171 &  $\;\;$97.5 & $0.96 \pm 0.02$ & 1,2,3 \\
Triangulum II   & -1.60 &      119.4 & $\;\;$30.0 & $6.7 \cdot 10^2$  & $\;\,$2.89 & 0.36 & 0.52 & 0.79 & 215 & 320 &  $\;\;$14.4 & $0.99 \pm 0.09$ & 5 \\
Tucana III      & -1.30 &      306.0 & $\;\;$22.9 & $5.1 \cdot 10^2$  & $\;\,$7.25 & 0.25 & 0.46 & 0.78 & 240 & 157 &  $\;\;$42.8 & $0.95 \pm 0.07$ & 20 \\
Tucana V        & -1.60 & $\;\;$60.0 & $\;\;$55.0 & $6.7 \cdot 10^2$  & $\;\,$2.55 & 0.48 & 0.61 & 0.79 & 140 & 227 &  $\;\;\;\;$1.3 & $1.00 \pm 0.12$ & 11 \\
Virgo I         & -0.80 & $\;\;$90.0 & $\;\;$87.0 & $3.2 \cdot 10^2$  & $\;\,$7.50 & 0.61 & 0.68 & 0.79 & 135 &  74 &  $\;\;$28.5 & $0.80 \pm 0.17$ & 21 \\
Willman 1       & -2.53 &      150.6 & $\;\;$38.0 & $1.6 \cdot 10^3$  & $\;\,$7.77 & 0.36 & 0.55 & 0.79 & 135 & 420 &  $\;\;$29.7 & $1.04 \pm 0.10$ & 5 \\
\hline
\end{tabular}}

\begin{flushleft}
References for distances, structural and photometric parameters: 1) \citet{baumgardtetal2019a}, 2) \citet{baumgardtetal2020}, 3) \citet{baumgardtvasiliev2021}, 4) \citet{longeardetal2019}, 
 5) \citet{munozetal2018}, 6) \citet{balbinotetal2013}, 7) \citet{walshetal2008},  8)~\citet{connetal2018}, 9)~\citet{connetal2018b}, 10) \citet{laevensetal2015}, 11) \citet{koposovetal2015}, 
 12)  \citet{drlicawagneretal2015}, 13)~\citet{martinez2019}, 14) \citet{jerjenetal2018}, 15) \citet{kimjerjen2015c}, 16) \citet{kimetal2015}, 17) \citet{kimetal2016}, 
  18)~\citet{garofaloetal2021}, 19) \citet{bechtoletal2015}, 20)~\citet{mutlupakdiletal2018}, 21) \citet{hommaetal2016}
\end{flushleft}
\end{table*}

\subsection{Globular clusters}

We first discuss our results for globular clusters. Fig.~\ref{fig:mseggc} depicts the $R_{H,Bright}/R_{H,Faint}$ ratios 
as a function of the dynamical cluster ages defined as the ratio of the physical age $T_{Age}$ of a cluster over its present-day relaxation time $T_{RH}$. Clusters with small values
of $T_{Age}/T_{RH}$, i.e. relaxation time equal to or larger than their age should not be strongly evolved dynamically and the internal distribution of stars should still reflect the initial distribution,
while clusters with large $T_{Age}/T_{RH}$ values on the other hand should be highly evolved and have developed strong mass segregation between their stars.

The red circles in Fig.~\ref{fig:mseggc} show the distribution of observed Milky Way globular clusters. It can be seen that all systems with large relaxation times have $R_{H,Bright}/R_{H,Faint}$ ratios very close to unity,
showing that these clusters are (nearly) unsegregated.  For increasing $T_{Age}/T_{RH}$ the amount of mass segregation increases and the most dynamically evolved clusters in our
sample have $T_{Age}/T_{RH}$ ratios around 0.7. These clusters also have $P$ values from the KS test that are close to 100\%, showing a very secure detection of mass segregation in these clusters.
The grey shaded region and black line in Fig.~\ref{fig:mseggc} depict the mass segregation results that we get from an analysis of the $N$-body simulations.
It can be seen that the amount of mass segregation in observed globular clusters is in full agreement with the observed amount of mass segregation in the simulated clusters. Since the simulated
clusters started without mass segregation, we take this 
as clear evidence that the globular clusters that we have analysed here also started without primordial mass segregation among their low-mass stars. Given that the clusters analysed span about two orders of
magnitude in mass and a factor 30 in Galactocentric distance, there is no reason to assume that our results can not be generalised to the other globular clusters as well.

\subsection{Star clusters/Ultra-faint dwarf galaxies}

We next analyse the ultra-faint dwarf galaxy candidates. The determination of mass segregation for them is done in exactly the same way as for the globular clusters.
Fig.~4 and Table~3 show the distribution of $R_{H,Bright}/R_{H,Faint}$ ratios that we obtain this way. Since the  ultra-faint dwarf galaxy candidates have generally fewer stars than
globular clusters, we obtain larger error bars on these ratios. Hence the $R_{H,Bright}/R_{H,Faint}$ ratios alone are not always sufficient to safely classify an object, so our final classification
also includes literature data whenever available. Before studying the full distribution, we therefore first discuss the available observational data of each individual object to derive a final
classification.

\subsubsection{Balbinot 1}

\citet{balbinotetal2013} discovered Balbinot~1 in Sloan Digital Sky Survey (SDSS) data and studied the system in more detail using CFHT/MegaCam data. They found a half-light radius of $r_h=7.2$ pc,
an age of about 11.7 Gyr
and an absolute luminosity of $M_V=-1.21 \pm 0.66$. The latter quantity is within the range of values measured for Galactic open clusters. Not much else is known about the system. Using the CFHT/MegaCam 
from \citet{balbinotetal2013}, we obtain $R_{H,Bright}/R_{H,Faint}=0.68 \pm 0.18$ for Balbinot~1 and a highly significant detection of mass segregation with $P_{Mseg}=99.8\%$. We therefore 
conclude that Balbinot~1 is a star cluster.

\subsubsection{Bo{\"o}tes II}

Bo{\"o}tes II was discovered in SDSS data by \cite{walshetal2007}, who speculated that it was a dwarf galaxy based on its large half-light radius (about 70 pc). Spectroscopic observations by
\citet{kochrich2014} and \cite{jietal2016} later found a very low metal abundance and a low fraction of neutron capture elements, in agreement with the assumption that Bo{\"o}tes II is a dwarf galaxy.
Our results are also compatible with this assumption since we find $R_{H,Bright}/R_{H,Faint}=0.95 \pm 0.07$, i.e. no significant detection of mass segregation. We conclude that Bo{\"o}tes II is a dwarf galaxy.

\subsubsection{Cetus II}

Cetus II was discovered by \citet{drlicawagneretal2015} who determined a heliocentric distance of 30~kpc and a compact size of $r_h=17$ pc. The parameters of Cetus~II were later revised by
\citet{connetal2018} using deep Gemini GMOS-S g, r photometry, who failed to find a stellar overdensity in Cetus II and, 
given the similarity in distance and metallicity, suggested that Cetus II is made up of stars from the Sagittarius dwarf galaxy. We also see no density concentration in our data, 
however the HST photometry we use covers a too small field of 
view to be conclusive. We obtain $R_{H,Bright}/R_{H,Faint}=1.12 \pm 0.20$ for Cetus II, compatible with no mass segregation and the system being either a dwarf galaxy or a tidal stream. We can
rule out a star cluster nature of Cetus II since the expected value of $R_{H,Bright}/R_{H,Faint}$ would then be about 0.70 to 0.80, depending on the fraction of the system that is 
covered by our photometry. We classify the nature of Cetus II 
as inconclusive. Radial velocities for some of the member stars might help to decide if Cetus II is a tidal tail of the Sagittarius galaxy or a separate stellar system.

\subsubsection{DES 1}

DES~1 might be a satellite of the Small Magellanic Cloud \citep{connetal2018} but its status as a dwarf galaxy or star cluster has so far not been established. 
\citet{connetal2018} found a $P_{Mseg}=14.8\%$ chance that the DES~1 is not mass segregated. From our analysis of the stellar distribution
in their ground-based Gemini/GMOS-S g, r data we find a similar, albeit somewhat lower chance of $P_{Mseg}=4.7\%$ that the bright
and faint stars follow the same distribution. The low probabilites together with the compact size of DES~1 therefore argue for the system being a star cluster.

\subsubsection{Draco II}

Draco II was discovered by \citet{laevensetal2015} in the Pan-STARRS1 3$\pi$ survey. According to \citet{longeardetal2018} it
is an old, metal-poor stellar system at a distance of about 21 kpc. \citet{martinetal2016} and \citet{longeardetal2018} carried out spectroscopic observations of Draco II but 
obtained only upper limits on the velocity dispersion.  \citet{longeardetal2018} also failed to find significant mass segregation in the analysis of their CFHT/MegaCam data.
We obtain $R_{H,Bright}/R_{H,Faint}=0.78 \pm 0.07$, clearly indicating mass segregation and in full agreement with the expected value for a dark matter free star cluster of
the mass and size of Draco~II. The difference to \citet{longeardetal2018} could be due to our deeper HST photometry which reaches nearly 2 magnitudes further below
the main-sequence turnover than their data.  We conclude that Draco II is star cluster.

\subsubsection{Eridanus III}

Eridanus III was discovered along with 8 other Milky Way satellite candidates by \citet{koposovetal2015} in Dark Energy Survey (DES) data. \citet{connetal2018} 
obtained \mbox{Gemini/GMOS-S}~g,~r photometry and derived a Galactocentric distance of $91 \pm 4$
kpc and an absolute luminosity of $M_V = -2.07 \pm 0.50$ for Eridanus III. From their data they did not find evidence for mass segregation in the 
system. Nevertheless, they classified Eridanus III as a star cluster mostly due to its compact size and low luminosity, which are similar to that of other 
known halo star clusters. Our results support this conclusion since we obtain a significant detection
of mass segregation and a mass segregation ratio of $0.84 \pm 0.09$ in full agreement with the expected one for a star cluster with the relaxation time of Eridanus III. 
The difference between their results and ours could again be due to the fainter magnitudes that we reach with our HST data, which allow us to probe less massive 
stars that show a larger degree of mass segregation.

\subsubsection{Grus I}

Like Eridanus III, Grus I was discovered by \citet{koposovetal2015} in Dark Energy Survey (DES) data. \citet{walkeretal2016} obtained Magellan/M2FS spectra of 7 likely members but
failed to find a velocity or metallicity spread among them. \citet{jietal2019} obtained high-resolution spectra for two member stars which still showed no metallicity difference
but revealed that both stars were deficient in neutron-capture elements. Since such an abundance pattern is rare among globular cluster stars, they concluded that Grus I is
a dwarf galaxy.  Further evidence for a dwarf galaxy nature of Grus I was presented by \citet{cantuetal2021}, who derived a half-mass radius of 150 pc for Grus I, 
much larger than expected for a star cluster. Furthermore \citet{jerjenetal2018} identified two stellar overdensities in Grus I, one of which is covered
by our HST photometry. We derive mass segregation parameters of $0.87 \pm 0.08$ for Grus I using the centre position given by \citet{jerjenetal2018} which
is mid-way between the two density peaks they identified. Adopting the density peak covered by the HST photometry as centre of Grus~I, this value changes 
to $0.91 \pm 0.08$. Both values could indicate that Grus I is mass segregated, something that is at variance with the large relaxation time of the system, which
is larger than a Hubble time even if Grus I would be free of dark matter. A possible explanation could be that Grus I is a dwarf galaxy with one or two small star clusters orbiting inside it. 

\subsubsection{Grus II}

Grus II was discovered by \citet{drlicawagneretal2015} in second year DES data. They estimated a half-light radius of 6 arcmin and a distance of 55 kpc to the system. The half-light radius would correspond 
to a physical size of about 95~pc, making it unlikely that Grus II is a star cluster. For the stars in our HST field, we find a half-number radius of 96 arcsec, corresponding to a lower limit of 25 pc 
on the half-light radius, in agreement with the size estimated by \citet{drlicawagneretal2015}. We also do not find significant mass segregation among the member stars ($P_{Mseg}=8.9\%$).
Grus II is therefore most likely a dwarf galaxy. We finally note that \citet{simonetal2020} recently obtained spectroscopy for three member stars but failed to find
a significant velocity dispersion or metallicity spread for them. Additional radial velocities might be necessary to finally settle the nature of Grus II.

\subsubsection{Horologium I}

Not much is known about the nature of Horologium I. Based on five member stars, \citet{koposovetal2015b} measured a velocity dispersion of $4.9^{+2.8}_{-0.9}$ km/sec for Horologium I. From the velocity dispersion they also derived a mass-to-light ratio of $M/L \sim 600$. However, given the small number of stars and the fact that the
radial velocities were taken at a single epoch, there is the possibility that their velocity dispersion is influenced by binary stars or non-members.
\citet{nagasawaetal2018} measured the abundances of three stars in the system and found a mean metallicity of [Fe/H]=-2.6
and very similar abundance patters between these stars. The low metallicity is more compatible with a dwarf galaxy but would not exclude a star cluster either. We find no significant mass segregation between the
stars ($P_{Mseg}=9.4\%$) and therefore conclude that Horologium I is most likely a dwarf galaxy.

\subsubsection{Horologium II}

There is some confusion about the correct density centre of Horologium II. \citet{kimjerjen2015c} give the position of the density centre as RA=49.13375, DEC=-50.01806, while \citet{munozetal2018} find 
RA=49.1077, DEC=-50.0486 with significant errors in both right ascension and declination. The stars selected as members in our HST data show a clear concentration around RA=49.138, 
DEC=-50.008 and we therefore adopt this position as the centre of Horologium~II. With that position, we obtain $R_{H,Bright}/R_{H,Faint}=1.05 \pm 0.11$, i.e. no evidence for mass segregation.
No further information is available for the system, however given the lack of mass segregation we conclude that Horologium~II is most likely a dwarf galaxy.

\subsubsection{Kim 2, Koposov 1, Koposov 2}

Our analysis of these three systems is based on the ground-based CFHT/Megacam data of \citet{munozetal2018}. Since this data is less deep  than the HST data that we use for most
of the other star clusters and dwarf galaxies, we can only analyse about $\sim$100 stars in each system. Nevertheless we get relatively strong detections of mass segregation for each system. Our detection of mass segregation in Kim~2 agrees with the results of \citet{kimetal2015}. Since the
compact sizes of these systems also argue in favour of them being star clusters, we conclude that the three systems are most likely star clusters.

\subsubsection{Pegasus III}

The most detailed study of Pegasus III was done by \citet{kimetal2016}, who obtained Keck/DEIMOS spectroscopy and Magellan/IMACS photometry of Pegasus III. They derived a large
half-light radius of 54 pc and a significant elongation of the system, something that we also see in our HST photometry. Based on eight stars,
they also determined a low metallicity and a stellar velocity dispersion of $\sigma=5.4_{-2.5}^{+3.0}$ km/sec, indicating a strongly dark matter dominated system.
However, \citet{kimetal2016} also
noticed that their solution for the velocity dispersion would drop to zero with the exclusion of the star with the most discrepant radial velocity. \citet{garofaloetal2021} obtained
$B,V$ time series photometry of Pegasus III and derived a 20\% smaller distance to the system. Their photometry also indicated a significant metallicity spread in the system, which
again would speak for a dwarf galaxy nature of Pegasus III. Our results are compatible with this conclusion since we find no evidence for mass segregation of Pegasus III. Taking all evidence
together, it seems most likely that Pegasus III is a dwarf galaxy, although a tidally disrupted star cluster might still be a possible solution. Additional spectroscopy should
help settle the question.

\subsubsection{Phoenix II and Pictoris I}

Phoenix II and Pictoris I were discovered by \citet{bechtoletal2015} and \citet{koposovetal2015} in Dark Energy Survey (DES) data. No velocity or metallicity information is available for
these systems. The structural parameters that have been determined for both systems by \citet{mutlupakdiletal2018} and \citet{munozetal2018} make them larger than normal star
clusters and are more typical for dwarf galaxies. Since we also do not detect significant mass segregation in both systems we conclude that they are most
likely dwarf galaxies, although given the large relaxation times, a star cluster nature is not completely ruled out either by our data.

\subsubsection{Reticulum II}

A dwarf galaxy nature of Reticulum II seems firmly established by its large velocity dispersion, which is far larger than expected for a purely stellar system \citep{walkeretal2015,simonetal2015,minoretal2019}
as well as the chemical abundance patterns found in its member stars by \citet{roedereretal2016b}, which closely resemble those of other confirmed dwarf galaxies.
Our mass segregation data is compatible with this conclusion, although, due to the large relaxation time which is larger than a Hubble time even in case of a purely stellar system, we would 
not be able to distinguish between a star cluster or a dwarf galaxy.

\subsubsection{Reticulum III}

No velocity or metallicity data is available for Reticulum III which would allow a classification of the system. We obtain a mass segregation ratio of $R_{H,Bright}/R_{H,Faint} = 1.20 \pm 0.10$, which is compatible
with no mass segregation. Given the lar lack of mass segregation and the large size of the system, we conclude that Reticulum III is a dwarf galaxy.

\subsubsection{Sagittarius II}

Sagittarius II was discovered as a stellar overdensity by \citet{laevensetal2015} in Pan-STARRS~1 data. \citet{longeardetal2020} and \citet{longeardetal2021} obtained stellar radial velocities and metallicity estimates
for about 25 member stars of Sagittarius~II from which \citet{longeardetal2021} determined a velocity dispersion of about $1.7 \pm 0.5$ km/sec. Such a velocity dispersion would be significantly higher
than the one expected for a stellar system with the size and mass of Sagittarius~II, since, using the $N$-body models of \citet{baumgardt2017}, we predict a velocity dispersion of only about 
0.5 km/sec for Sagittarius~II. The measured velocity dispersion of Sagittarius~II might however be inflated by binaries. From our mass segregation test we obtain a mass segregation parameter of
$0.96 \pm 0.02$, in full agreement with the expected value for a star cluster. We therefore conclude that Sagittarius~II is a star cluster. This classification is in agreement with the fact that
 \citet{longeardetal2021} find no metallicity spread among the member stars of Sagittarius~II.

\subsubsection{Segue 1}

Segue~1 seems almost certainly a dwarf galaxy based on its large velocity dispersion \citep{simonetal2011} as well as the large metallicity spread seen among its giant stars \citep{frebeletal2014}.
We obtain a mass segregation ratio of $0.93 \pm 0.06$, fully compatible with an unsegregated dwarf galaxy and significantly higher than expected for a star cluster of the size of Segue~1 (about 0.80). 
We therefore conclude that Segue~1 is a dwarf galaxy.

\subsubsection{Segue 2}

\citet{kirbyetal2013} identified Segue 2 as a dwarf galaxy based on the wide spread in metallicity that they found among the member stars, even though they were not able to resolve its velocity dispersion.
Our mass segregation data is compatible with this assumption since we do not detect significant mass segregation among the member stars ($P_{Mseg}=4.5\%$). We therefore classify Segue~2 as a dwarf galaxy. 

\subsubsection{Segue 3}

\citet{fadelyetal2011} performed the first detailed kinematic and photometric study of Segue~3. They derived an age of 12 Gyr and found no significant metallicity spread among the member stars. Since they were 
also not able to resolve the velocity dispersion, they concluded that Segue 3 is an old star cluster. The age of Segue~3 was later revised downwards to 3.2 Gyr by \citet{ortolanietal2013} and 2.5 Gyr by
\citet{hughesetal2017}. We find a moderately significant detection of mass segregation in Segue 3 ($P_{Mseg}=75.9\%$). Together with the young age and its location in a mass vs. size relation 
we therefore conclude that Segue~3 is a star cluster.

\subsubsection{Triangulum II}

\citet{martinetal2016b} obtained Keck/DEIMOS spectroscopy for 13 member stars of Triangulum II and found a high central velocity dispersion of $\sigma = 4.4^{+2.8}_{-2.0}$ km/sec. They also found that the
velocity dispersion increased further outwards, indicating either a strongly dark matter dominated system or the ongoing tidal disruption of the system. \citet{kirbyetal2015b} confirmed the
high central velocity dispersion, and also found a metallicity spread of $\Delta$ [Fe/H]=0.8 dex among the member stars. However \citet{kirbyetal2017} obtained additional velocities for 13 member stars and obtained only an
upper limit of 3.4 km/sec for the velocity dispersion. In addition, they also noted that the metallicity spread detected among the member stars hinges on the membership of the two most metal-rich stars,
leaving open the possibility that Triangulum II is a star cluster. We obtain a mass segregation parameter $R_{H,Bright}/R_{H,Faint} = 0.99 \pm 0.09$, in good agreement with the assumption that Triangulum II
is unsegregated. As a star cluster we would expect to find a value around 0.80 due to the low relaxation time of Triangulum II. We therefore conclude that the system is a dwarf galaxy.

\subsubsection{Tucana III}

Tucana III is unique among the known ultra-faint dwarf galaxy candidates of the Milky Way since it has confirmed tidal tails \citep{lietal2018} indicating ongoing tidal disruption of the system. 
However the nature of Tucana III is still not certain.
\citet{simonetal2017} presented radial velocities for 26 member stars of Tucana III, from which they were only able to derive an upper limit of 1.5 km/sec for the velocity dispersion.
They also did not detect a significant metallicity spread among those stars. \citet{marshalletal2019} presented high-resolution spectroscopy for four member stars and found a moderate enrichment
in $r$-process elements. They argued that the abundance pattern suggest a dwarf galaxy nature for Tucana III. Our data is inconclusive, since we find a mass segregation ratio of
$0.95 \pm 0.07$, compatible with both a mass segregated star cluster or a dwarf galaxy without mass segregation. However, based on
the results of \citet{marshalletal2019}, we assume that Tucana~III is more likely a dwarf galaxy.

\subsubsection{Tucana V}

\citet{simonetal2020} obtained Magellan/IMACS spectroscopy for 3 stars in Tucana V but found neither a significant metallicity spread nor a velocity dispersion of the system.
The photometry by \citet{connetal2018} was equally inconclusive since they found that Tucana V is compatible with the location of dwarf galaxies in a luminosity-metallicity plane,
but compatible with a star cluster in a size-luminosity plane. We find no significant mass segregation for Tucana V ($P_{Mseg}=1.3\%$) while as star cluster it should be strongly mass segregated and 
therefore conclude that Tucana V is a dwarf galaxy.

\subsubsection{Virgo I}

Virgo I was discovered by \citet{hommaetal2016} in data from the  Subaru/Hyper Suprime-Cam survey. Nothing much is known about the system. Our HST photometry shows a clear
concentration of member stars around RA=180.035, DEC=-0.690, slightly different to the centre position given by \citet{hommaetal2016}. Adopting this position as density centre,
we find a mass segregation parameter $0.80 \pm 0.17$, compatible with either a mass segregated or an unsegregated system. We notice however that the spatial distribution
of stars in our HST field appears very elongated. Furthermore the absolute luminosity and distance given by \citet{hommaetal2016} would result in a tidal radius for
Virgo I of about 45~pc, not much larger than the quoted half-mass radius of 38~pc. Virgo I might therefore not be a bound system, although with the data at hand
we can't distinguish between a tidal tail of a dwarf galaxy or that of a star cluster.

\subsubsection{Willman 1}

Willman 1 was discovered by \citet{willmanetal2005} in SDSS data. \citet{willmanetal2011} obtained  Keck/DEIMOS spectroscopy for 45 likely member stars,
and, while they were not able to detect a velocity dispersion for the system, they found a significant metallicity spread between two member stars. They therefore concluded that
Willman 1 is a dwarf galaxy. We find $R_{H,Bright}/R_{H,Faint} = 1.05 \pm 0.08$, indicating that Willman~1 is unsegregated. The expected value for a star cluster would only be
around 0.90, which is significantly below the observed ratio. We therefore conclude that Willman~1 is a dwarf galaxy. 

\subsection{Final results}

\begin{figure*}
\begin{center}
\includegraphics[width=0.95\textwidth]{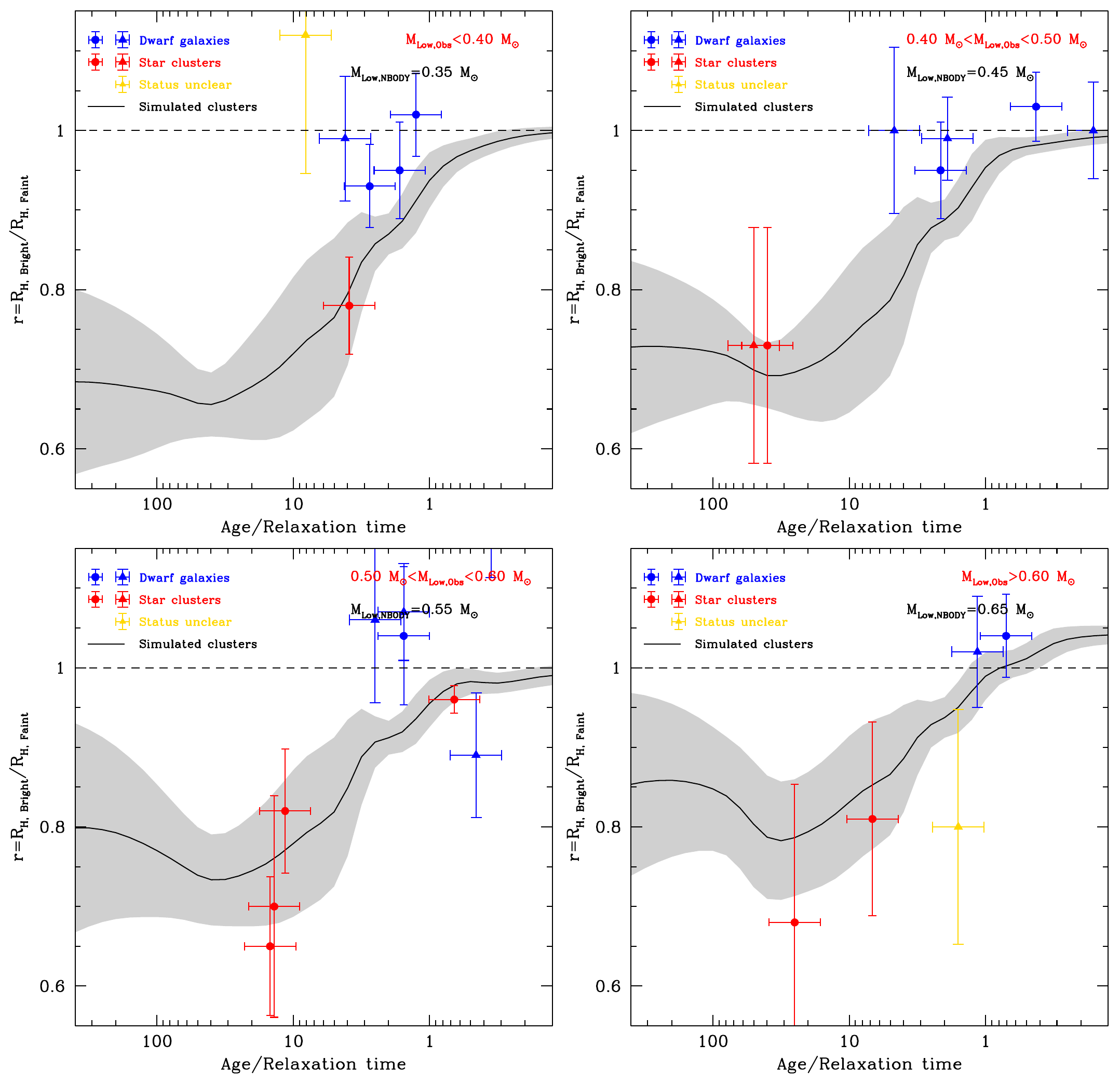}
\caption{Same as Fig.~\ref{fig:mseggc} for the dwarf galaxies and star clusters. Red, blue and yellow symbols show our final classification based on the measured  $R_{H,Bright}/R_{H,Faint}$ ratios as well as literature
data on metallicities and velocity dispersions. Circles/triangles mark star clusters and dwarf galaxies without/with literature information on their nature. There is a clear dichotomy in the mass segregation distribution. Star clusters follow the same trend between mass segregation and their relaxation times as seen for globular clusters and simulated star
clusters while the dwarf galaxies are all compatible with no mass segregation.}
\label{fig:msegufd}
\end{center}
\end{figure*}

Fig.~\ref{fig:msegufd} depicts the final distribution of the mass segregation ratios as a function of the relaxation time for all dwarf galaxy candidates. Objects
that we classify as star clusters are shown in red, dwarf galaxies are shown in blue and objects with an unclear status in yellow.
The only two objects that we cannot classify are Cetus II and Virgo I, which are most likely tidal streams.
Our final classification takes the observed mass segregation ratios
as well as literature  results on metallicity and velocity dispersion into account. In order to allow a comparison of our classification with existing literature, we show 
with triangles objects that have either a resolved velocity dispersion, a significant abundance spread among the member stars or an heavy element abundance pattern that makes them
likely dwarf galaxies. Objects without such signs are shown by circles. 

It can be seen that our mass segregation results provide a clear separation between dwarf galaxies and star clusters since most objects either have no obvious mass segregation
or are clearly mass segregated. In addition the classification based on our mass segregation ratio generally agrees with the literature classification based on metallicity or
velocity dispersion (blue and red triangles). The only exception is Grus I, where our mass segregation classification could be influenced by the substructure found by \citet{jerjenetal2018}.

\begin{figure*}
\begin{center}
\includegraphics[width=1.00\textwidth]{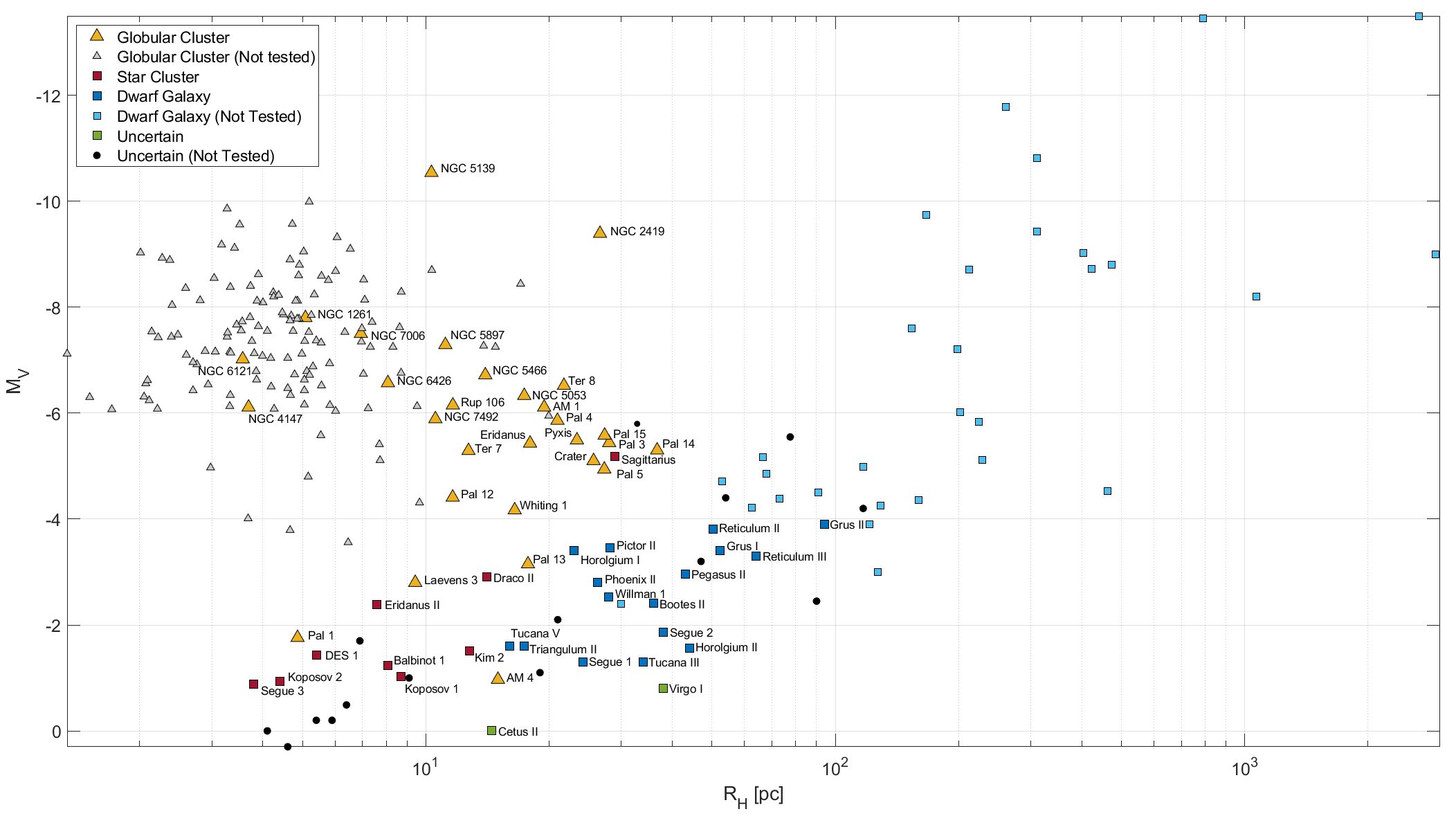}
\caption{Half-mass radius vs. absolute magnitude for all dwarf galaxy candidates and globular clusters. Systems studied in this work are marked by their names. Systems classified as star clusters are
shown by red squares, those classified as dwarf galaxies by blue squares and those without a final classification by green squares. We also performed a literature search to classify the systems that
were not studied in detail in this paper. Systems that have no classification in the literature are shown by black circles.}
\label{fig:massvssize}
\end{center}
\end{figure*}

Fig.~\ref{fig:massvssize} finally shows the distribution of known Galactic globular clusters and dwarf galaxies in a mass vs. size plane. Globular clusters studied in this paper are
marked by yellow triangles, star clusters by red squares and dwarf galaxies are shown as blue squares and objects without a classification as black circles.
For systems not studied in this work we have searched the literature to classify them into dwarf galaxies or star clusters. For most objects this classification was
relatively straightforward since they are relatively large objects with sizes of 100~pc or more that show clear evidence of being dwarf galaxies.

There is a relatively clear division between dwarf galaxies and star clusters in Fig.~\ref{fig:massvssize}. All known systems with sizes $r_h>20$ pc are dwarf galaxies, with
the exception of a few globular clusters. These are however significantly more luminous than the dwarf galaxies of the same size and occupy a region in parameter
space quite distinct from the dwarf galaxies. The possible unbound systems Cetus II and Virgo I are significantly less luminous than dwarf galaxies of comparable size,
which again shows that they could be different from the other dwarf galaxies. Interestingly there
is no confirmed dwarf galaxy with size of less than 20 pc. This could be a true limit meaning lower mass halos don't exist or were not able to form any stars.
Alternatively, given that about half of all compact and low-mass objects in Fig.~\ref{fig:massvssize} have no classification yet,  we might have simply
not found more compact dwarf galaxies. 

\begin{figure*}
\begin{center}
\includegraphics[width=0.95\textwidth]{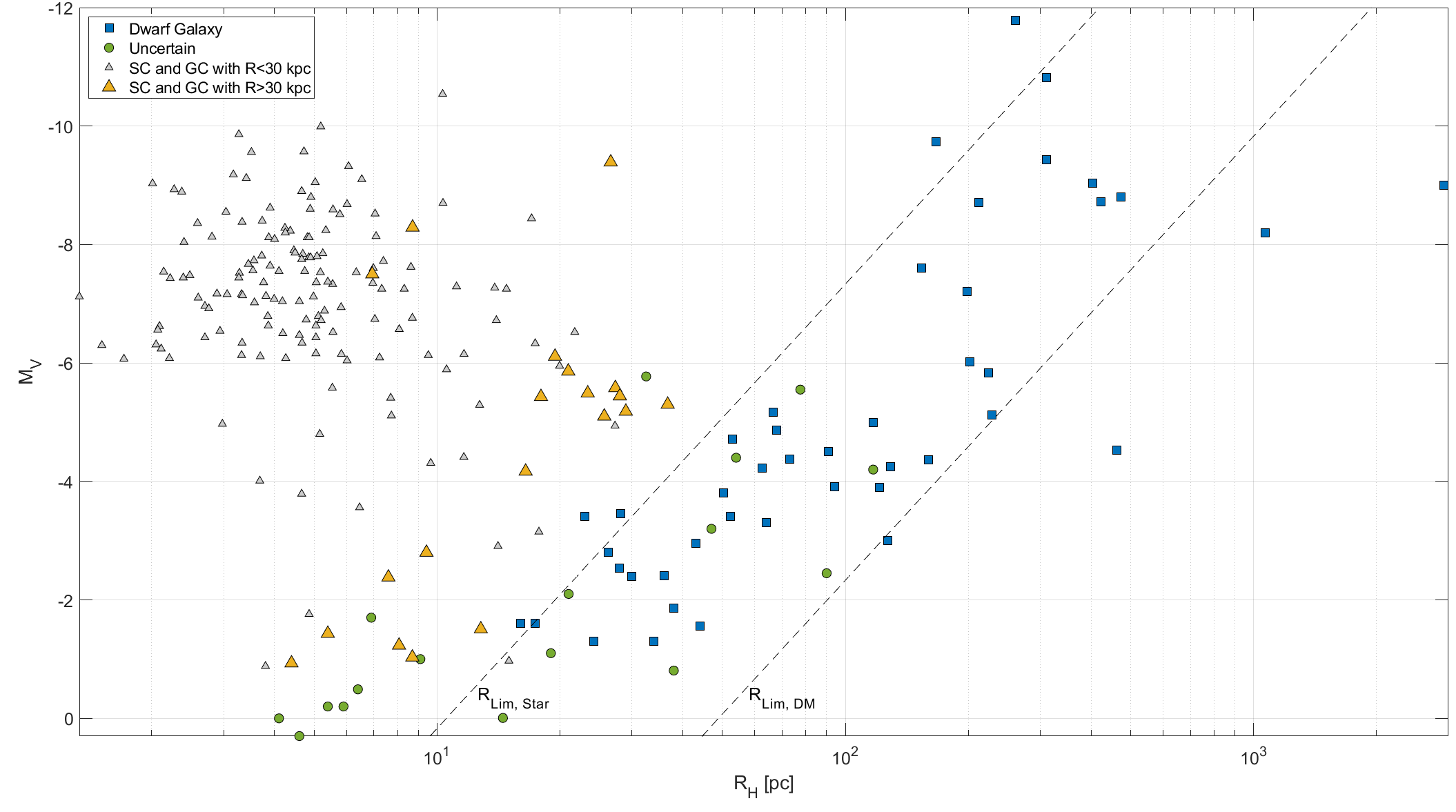}
\caption{Similar to Fig.~\ref{fig:massvssize} but now also showing limits where a stellar system should be tidally disrupted by the gravitational field of the Milky Way. The line $R_{Lim, Star}$ shows
the tidal limit for a purely stellar systems while the line $R_{Lim, DM}$ is for a system with $100 \times$ as much dark matter as luminous matter. Both lines are for a Galactocentric distance
of 80~kpc. All star clusters are to the left of the $R_H=R_{Lim, Star}$ line and most dwarf galaxies are to the left of the $R_H=R_{Lim, DM}$ line, implying that
most are bound systems. Unclassified systems with $R_H>R_{Lim, Star}$ can't be bound star clusters and are likely dwarf galaxies or undergoing tidal disruption.}
\label{fig:massvssize2}
\end{center}
\end{figure*}

Fig.~\ref{fig:massvssize2} compares the location of all systems in a size vs. luminosity plane with the tidal radius of the systems due to the gravitational field of the Milky Way
added. For a system of mass $M$ in circular orbit with radius $R_G$ in a logarithmic potential with circular velocity $V_C$, the tidal radius is given by
\begin{equation}
 R_{Tide}= \left(\frac{M}{2 V_C^2} \right)^{1/3}  R_G^{2/3}
\end{equation}
The tidal radius describes the maximum radius out to which a system can keep stars bound. Tides will start to remove the outer stars well before the half-mass radius
of a system equals its tidal radius, leading to mass loss and a further shrinking of the tidal radius. Dissolution should therefore already set in well before the
half-mass radius of a dwarf galaxy or star cluster approaches its tidal radius. Assuming a maximum ratio $R_H/R_{Tide}=0.33$ necessary for stability and that the projected half-light radius that observers
measure is 0.75 times the 3D one, we obtain $R_{Lim} = 0.25$ $R_{Tide}$ as the maximum radius that any system can have before tidal disruption happens.
The two dashed lines in Fig.~\ref{fig:massvssize2} show this relation for the case of a pure stellar system with $M_{Tot}=M_{Lum}$ and a dark matter dominated system
with $M_{Tot}=100 M_{Lum}$. For both lines we assume a  Galactocentric distance of 80~kpc and a circular velocity of 170 km/sec \citep{deasonetal2021} at this distance.
We restrict ourselves to objects with $R_{GC}>30$ kpc in this plot, and show globular clusters at smaller distances with a different symbol.

It can be seen that all objects classified as star/globular clusters stay to the left of the tidal limit line for purely stellar systems and should therefore be stable to tidal forces even in the
absence of dark matter, confirming our classification. Objects that have no classification so far occupy either side of the $R_{Lim, Star}$ line. As star clusters with $r_{h,p}>R_{Lim, Star}$
should undergo quick dissolution, the unclassified objects with $r_{h,p}>R_{Lim, Star}$ are most likely dark matter dominated dwarf galaxies or star clusters in the process of disruption.
Given that we find only very few confirmed dwarf galaxies with $r_{h,p}<R_{Lim, Star}$, it seems likely
that the unclassified objects with $r_h<10$~pc are star clusters. Most dwarf galaxies have $r_{h,p}<R_{Lim, DM}$, meaning they
should be stable as long as they are strongly dark matter dominated. The only systems to the right of this line are Antila~2 and Crater II, for which \citet{jietal2021} detected signs 
of ongoing tidal disruption, and Bo{\"o}tes IV for which \citet{hommaetal2019} found a large intrinsic ellipticity of $\epsilon=0.64$ which could indicate tidal interaction
and distortion. Furthermore Bo{\"o}tes IV is at a Galactocentric distance of $\sim 210$ kpc and has a relatively large error bar on its half-mass radius. Hence it might 
have $r_{h,p}<R_{Lim, DM}$ and be bound.

Since the tidal radius of a system increases with its mass to the power of $1/3$, the derived limits can be turned into densities. We find that star clusters in the outer halo have densities 
(within their half-light radii) between $3 \cdot 10^{-2}$ M$_\odot < \rho_h < 1$ M$_\odot$ while dwarf galaxies have (stellar) densities of $10^{-3}$ M$_\odot < \rho_h < 3 \cdot 10^{-2}$ M$_\odot$, 
about a factor 30 smaller than star clusters. Adding the dark matter to these numbers would make the densities of the dwarf galaxies roughly comparable to those of the star clusters.

Dwarf galaxies and star clusters in the halo of the Milky Way are subject to tidal forces from the Milky Way, which can remove mass from the system. If the mass loss is strong enough this could change the expected 
mass segregation signal and therefore influence our classifications. However, our classifications are unlikely to be strongly influenced by tidal effects since, with a few exceptions like Tuc~III, most 
dwarf galaxies do not show obvious tidal tails \citep{simon2019}. In addition, \citet{hammeretal2021} recently found from an analysis of the orbital motions of dwarf galaxies that most of them
are probably on their first passage of the Milky Way, again arguing against significant tidal stripping of most them. Finally the lifetimes calculated by \citet{baumgardtetal2019a} significantly exceed 
a Hubble time for most globular clusters studied in our paper, again arguing against strong mass loss.

 
\section{Conclusions}

We have derived the amount of mass segregation in over 50 globular clusters and dwarf galaxy candidates by analysing their colour-magnitude diagrams based on deep \hst and ground-based photometry.
We find that dynamically young globular clusters (i.e. clusters with relaxation times of the order of or larger than their ages) have little to no mass segregation and that
the amount of mass segregation increases with increasing dynamical age. At each dynamical age the amount of mass segregation
is fully compatible with the one found in $N$-body simulations of initially unsegregated clusters. Our results therefore strongly indicate that globular clusters form without primordial 
mass segregation between their low-mass stars. This conclusion is strengthened by the fact that the investigated globular clusters cover the full mass range seen for globular
clusters (from $3 \cdot 10^2$ M$_\odot$ to $3 \cdot 10^6$~M$_\odot$), and nearly the full metallicity range and range of Galactocentric radii (6.7~kpc to 147~kpc).
The only alternative explanation would be if globular clusters form segregated but the initial mass segregation is 
erased by the primordial gas expulsion process \citep[e.g.][]{marksetal2008,haghietal2015}. However it is difficult to see how this process could almost completely erase the initial mass segregation
signature, so the assumption of no initial mass segregation is by far the more natural assumption.

The lack of mass segregation that we find in the least evolved clusters is in agreement with observations of young open clusters which also 
find the low-mass stars to be unsegregated \citep[e.g][]{parkeretal2012,parker2017}. Our study leaves open the question of primordial mass segregation of the more massive stars,
for which both hydrodynamical simulations and observations show that they form preferentially in the highest density regions of a star forming molecular cloud 
\citep[e.g][]{bonnelldavies1998,maschbergerclarke2011,dibhenning2019}. 

We find that the dwarf galaxy candidates fall mostly into two groups, star clusters that are mass segregated in the same way as globular clusters of the same dynamical age and
unsegregated systems. Most unsegregated systems have dynamical ages larger than unity, so some additional process must be invoked to prevent mass segregation in these
systems. The most natural mechanism is a significant amount of dark matter. This is corroborated by the fact that many unsegregated dwarf galaxy candidates also have
either large velocity dispersions, a significant spread in [Fe/H] or $r$-process metal abundances typical of dwarf galaxies. Combining our mass segregation data with 
published literature data, we are thus able to classify almost all investigated dwarf galaxy candidates as either star clusters or dwarf galaxies. The only exceptions 
are Cetus~II and Virgo~I, which could be tidally disrupting systems, but is not clear if they are former star clusters or dwarf galaxies.

We finally find that star clusters and dwarf galaxies occupy narrow strips in the size vs. mass plane. Star clusters in the outer halo have densities (within their half-light radii)
between 0.03~M$_\odot$/pc$^3 \lesssim \rho_h \lesssim $ 1~M$_\odot$/pc$^3$ while dwarf galaxies have (stellar) densities of 0.001 M$_\odot$/pc$^3 \lesssim \rho_h \lesssim$ 0.03~M$_\odot$/pc$^3$,
about a factor 30 smaller
than the star clusters. The dividing line between both populations roughly corresponds to the limit where star clusters would be tidally disrupted by the tidal field of the Milky
Way. Dwarf galaxies seem to be limited in the same way if their dark matter is taken into account and all systems with stellar densities $\rho_h<3 \cdot 10^{-4}$ M$_\odot$/pc$^3$ 
show signs of ongoing tidal disruption. \citet{vandenbergh2008} suggested a division corresponding to $M \sim R_H^{5.7}$ (for constant mass-to-light ratio) between
globular clusters and dwarf galaxies, which seems too steep since many of the confirmed ultra-faint dwarf galaxies would end up on the star cluster side of this relation. Also the
commonly used division based on surface brightness (corresponding to a $M \sim R_H^2$ relation for constant mass-to-light ratio) seems to work less well since many star clusters would lie on the dwarf galaxy side.

\section*{Acknowledgments}

We thank an anonymous referee who helped improve the presentation of the paper.
This work is based on observations made with the NASA/ESA Hubble Space Telescope, obtained from the data archive at the Space Telescope Science Institute. STScI is operated by the Association of Universities for Research in Astronomy, Inc. under NASA contract NAS 5-26555.

\section*{Data Availability}

Data is available upon request.

\bibliographystyle{mn2e}
\bibliography{mybib}

\label{lastpage}

\end{document}